\documentclass[11pt]{article}

\usepackage{fullpage}
\usepackage{amssymb}
\usepackage{amsmath}
\usepackage{amsfonts}
\usepackage{amsthm}
\usepackage{url}
\usepackage{graphicx}
\usepackage{subfigure}
\usepackage{fancyhdr}
\usepackage{url}

\usepackage[utf8]{inputenc}
\usepackage[english]{babel}
\usepackage{hyperref}

\newcommand{\nc}{\newcommand}
\nc{\heading}[1]{\begin{center} \large \bf #1 \end{center}}

\newcommand{\oA}{\overline{A}}
\newcommand{\oB}{\overline{B}}
\newcommand{\oI}{\overline{I}}

\newcommand{\oR}{\overline{R}}

\newcommand{\Ex}{\mathsf{E}}

\theoremstyle{plain}

\usepackage{glossaries}
\makeglossaries
\newglossaryentry{latex}
{
    name=latex,
    description={Is a mark up language specially suited
    for scientific documents}
}
\newglossaryentry{maths}
{
    name=mathematics,
    description={Mathematics is what mathematicians do}
}

\begin{document}

%\copyrightnotice
%\lipsum[1-10]

\date{January 25, 2021}
\title{Stochastic Modeling of an Infectious Disease \\
 Part II: Simulation Experiments and Verification of the Analysis.
\footnote{This article was originally published on August 5, 2020 in the authored HP \cite{kobayashi:2020b}. Additional simulation results were reported in the keynote at ITC-32. For the slides and video, see \url{https://hp.hisashikobayashi.com/a-stochastic-model-of-an-infectious-disease/}.
} } 
\author{Hisashi Kobayashi\footnote{The Sherman Fairchild University Professor of Electrical Engineering and Computer Science, Emeritus, ~~
Email: Hisashi@Princeton.EDU,~~ Home page: \url{https.hp.HisashiKobayashi.com} Wipipedia: \url{https://en.wikipedia.org/wiki/Hisashi_Kobayashi}
} \\
   Dept. of  Electrical Engineering \\
   Princeton University \\
   Princeton, NJ 08544, U.S.A.}

\maketitle

\begin{center}\textbf{Abstract}\end{center}
In \cite{kobayashi:2020a}, we introduced a \emph{stochastic model} of an infectious disease, based on the \emph{birth-and-death-with-immigration} (BDI) process. The model can capture the essence of dynamics of an infection process.  The most significant finding is that the \emph{time-dependent} (or transient) probability distribution of the infected population size is \emph{negative binomial distributed} at all times with a very long tail distribution. Consequently, an epidemic pattern exhibits a much larger variation than has been considered by most modeling experts, including the epidemiology community.
  
In this report, Part II, simulation runs and their interpretations are presented to support the above finding.  The sample paths of different simulation runs exhibit indeed  enormous variations, which confirm our analysis. 
   
The epidemic pattern that a given environment (e.g., city or country) experiences is merely one \emph{sample path} out of infinitely many possible paths. The enormous variations we observe in simulation runs explain why some cities and countries experience much fewer infections and casualties than others. The size of the infected or dead populations could differ by factor of 100 or more even under identical statistical conditions.

There are two important implications of our findings.  First, it would be a futile effort to attempt to identify all possible causes or reasons why environments of similar situations differ so much in terms of epidemic patterns and the number of casualties.  Our stochastic model suggests that ``luck"  or ``chances" play a more significant role than most of us would be led to believe.  Second, we should be prepared for a worst possible scenario, which only a stochastic model can provide with some probabilistic qualification, rather than preparing for the future, based on a single prediction curve that a deterministic model can provide. 

For all these probabilistic arguments, however, our analysis also shows that the infection process will immediately start declining, once the exponential parameter $a=\lambda-\mu$ should turn negative.  This result will be discussed in full, by analysis \cite{kobayashi:2021a} 
and simulation \cite{kobayashi:2021b}, where we assume the infection rate parameter $\lambda$ and the recovery (which includes removal and death) parameter $\mu$ are arbitrary functions of time.  
  
\paragraph{\em Keywords:}
 Stochastic vs. deterministic models, Birth-and-death process with immigration (BDI), Event-driven simulation, Random number generator, 
 Sample paths of the processes $I(t)$, $R(t)$ and $R(t)$, The death process $D(t)$, Probability generating function,  Time-dependent PMF, Daily new infections, Percentile curves, Negative binomial distribution, Probability distribution with a long tail, Large coefficient of variation, Branching process, Disparity or large variation among different sample paths, Analogy to disparity in wealth distribution, Statistical fluctuation in the initial phase, Law of large numbers.
 
\tableofcontents

\section{Recapitulation of Part I}  \label{sec-Part_I}
Our model formulation began with the set of linear differential-difference equations (Part I, page 8, replicated below in (\ref{Diff_eqn_P_0(t)}) and (\ref{Diff_eqn_P_n(t)}) ) for the probability mass function (PMF)
\begin{align}
P_n(t) \triangleq\mbox{Pr}[I(t)=n], ~~n=0, 1, 2, \cdots, ~~t\geq 0,\label{PMF}
\end{align}
where $I(t)$ is the number of \emph{infected, hence infectious persons}, at time $t$.  We can express $I(t)$ as
\begin{align}
I(t)=I_0+ A(t)+B(t)-R(t),  \label{I(t)-def}
\end{align}
where $I_0$ is the initial value:
\begin{align}
I(0)=I_0,  \label{I_0}
\end{align} 
and
\begin{enumerate}
\item $A(t)$ is the \emph{cumulative count} of infected arrivals from the outside. We assume the arrival pattern is completely random, i.e., a Poisson process with rate $\nu$ [persons/day].

\item $B(t)$ is the \emph{cumulative count} of the infections that occur in the interval $[0, t]$. An infection occurs at the rate $\lambda$ [person/day/infectious person].

\item $R(t)$ is the \emph{cumulative count} of the infected persons who recover, are removed or die in $[0, t]$.  This event occurs at the rate $\mu$ [persons/day/infected person].
  
\end{enumerate}

The $P_n(t), n=0, 1, 2, \cdots$ should satisfy the following set of differential equations:
\begin{align}
\frac{dP_0(t)}{dt}&= - \nu P_0(t)+\mu P_1(t),  \label{Diff_eqn_P_0(t)}\\
\frac{dP_n(t)}{dt}&=[(n-1)\lambda+\nu]P_{n-1}(t)-[n(\lambda+\mu)+\nu]P_n(t)+(n+1)\mu P_{n+1}(t), ~~n=1, 2, 3. \cdots,\label{Diff_eqn_P_n(t)}
\end{align}
with the initial condition (\ref{I_0}).

In order to find solve the above differential equations, we use the \emph{probability generating function} (PGF) defined by
\begin{align}
G(z,t)\triangleq \Ex[z^{I(t)}]=\sum_{n=0}^\infty P_n(t)z^n,  \label{PGF}
\end{align}
The set of differential equations (\ref{Diff_eqn_P_0(t)}) and (\ref{Diff_eqn_P_n(t)}) are then transformed into one \emph{partial differential equation} (PDE):
\begin{align}
\frac{\partial G(z,t)}{\partial t}=(z-1)\left[(\lambda z-\mu)\frac{\partial G(z,t)}{\partial z}+\nu G(z,t)\right],\label{PDE-for-PGF}
\end{align}
with the condition
\begin{align}
G(z, 0)=z^{I_0}. \label{boundary-cond}
\end{align}

We apply Lagrange's method to the above PDE, obtaining
\begin{align}
G(z,t)=\left(\frac{a}{\lambda z-\mu-\lambda(z-1)e^{at}}\right)^r
\left(\frac{\lambda z-\mu-\mu(z-1)e^{at}}{\lambda z-\mu-\lambda (z-1)e^{at}}\right)^{I_0}, \label{Solution-PGF}
\end{align}
where
\begin{align}
a\triangleq\lambda-\mu, ~~\mbox{and}~~r\triangleq\frac{\nu}{\lambda}. \label{def-a}
\end{align}
If $I_0=0$,  (\ref{Solution-PGF}) reduces to 
\begin{align}
G(z,t)&=\left(\frac{a}{\lambda z-\mu-\lambda(z-1)e^{at}}\right)^r
=\left(\frac{1-\beta(t)}{1-\beta(t)z}\right)^r,  \label{46-Part-I}
\end{align}
where 
\begin{align}
\beta(t)\triangleq\frac{\lambda(e^{at}-1)}{\lambda e^{at}-\mu}.  \label{45-Part-I}
\end{align}

To obtain $\Ex[I(t)]=\oI(t)$, we use the formula $\Ex[I(t)]=\left.\frac{\partial G(z,t)}{\partial z}\right|_{z=1}$:
\begin{align}
\oI(t)=\frac{r\beta(t)}{1-\beta(t)}=\frac{\nu}{a}(e^{at}-1).\label{mean-I(t)}
\end{align}
It would be easy to observe that
\begin{itemize}
\item If $a>0$,  $\oI(t)$ grows exponentially to infinity, as $t\to\infty$; 
\item If $a<0$, it converges to $-\frac{\nu}{a}>0$.
\item If $a=0$, $\oI(t)=\nu t$ for all $t\geq 0$.  
\end{itemize}

Let $T$ [days] be the number of days that is required for $\oI(t)$ to double. Then from (\ref{mean-I(t)}), we find
\begin{align}
e^{aT}=2, ~~\mbox{or}~~aT=\ln 2=0.693.
\end{align}
If $a=0.2$ [day$^{-1}$] (as in the running example of Part I and the present paper), $\oI(t)$ doubles in every $T \approx 3.5$ [days], hence quadruples every week.\footnote{When the exponential growth rate in the environment of your interest differs from our example, you can still interpret our simulation results meaningfully. For example if $\oI(t)$ double every week in the environment of your interest, you can scale the time axis by factor of two, i.e., $t=0, 1, 2 \cdots, 50$ in our simulation plots, you assign different time scale, $t'=0, 2 ,4, \cdots, 25$.}

In order to obtain the PMFs $P_n(t)$, by referring to the definition of the PGF (\ref{PGF}), we differentiate  $G(z,t)$ (\ref{46-Part-I}) w.r.t. $z$, set $z=0$, divide the resulting expression by $n!$.  That is,
\begin{align}
P_0(t)&=G(0,t)=(1-\beta(t))^r\nonumber\\
P_1(t)&=\left.\frac{\partial G(z,t)}{\partial z}\right|_{z=0}=(1-\beta(t))^rr\beta(t)\nonumber\\
P_2(t)&=\left.\frac{1}{2!}\frac{\partial^2 G(z,t)}{\partial z^2}\right|_{z=0}=(1-\beta(t))^r\frac{(r+1)r}{2!}\beta(t)^2\nonumber\\
\vdots &  \nonumber\\
P_n(t)&=\left.\frac{1}{n!}\frac{\partial^n G(z,t)}{\partial z^n}\right|_{z=0}=(1-\beta(t))^r\frac{(n+r-1)\cdots(r+1)r}{n!}\beta(t)^n,  \label{P_n(t)-from-PGF}
\end{align}
which we can write as
\begin{equation}
 \fbox{
\begin{minipage}{9.5cm}
\[
P_n(t)=K(n,r)(1-\beta(t))^r \beta(t)^n, ~~n=0, 1, 2, \cdots,    
\] 
\end{minipage}
}  \label{PMF-I(t)}
\end{equation}
where $K(n,r)$ is the binomial coefficient:
\begin{align}
K(n,r)&=\frac{(n+r-1)(n+r-2)\cdots (r+1)r}{n!}\triangleq {n+r-1\choose n}.\label{Binomial-coeff}
\end{align}
The above distribution $P_n(t)$ takes the form of a \emph{Pascal distribution}, or \emph{negative binomial distribution} (NBD), often denoted as NB($k, q$), which is the probability of the number of failures $n$ in Bernoulli trials needed to achieve $k$ successes, where $q$ is the probability of failure per trial
\begin{align}
P^{\scriptscriptstyle NBD}_n=K(n,k) (1-q)^k q^n,~~n=0, 1, 2, \cdots.\label{PMF-NBD}
\end{align}
The PGF of this NBD is given by
\begin{align}
G_{\scriptscriptstyle NBD}(z)\triangleq\sum_{n=0}^\infty P^{\scriptscriptstyle NBD}_n z^n=\left(\frac{1-q}{1-qz}\right)^k.  \label{PGF-NBD}
\end{align}
Thus, by equating $k=r\triangleq\frac{\nu}{\lambda}$ and $q=\beta(t)$, we see from (\ref{46-Part-I}) that the BDI process $I(t)$ is negative binomial distributed according to NB($r, \beta(t)$) at given $t$.

We note the following properties of $K(n,r)$ (\ref{Binomial-coeff}) and the NBD (\ref{PMF-I(t)}): 
\begin{itemize}
\item[i] $K(0,r)=1$, thus $P_0(t)=(1-\beta(t))^r$ for all $t\geq 0$.
\item[ii] $K(1,r)=r$, thus $P_1(t)=r(1-\beta(t))^r\beta(t)$. Therefore,  $P_1(t)=r\beta(t)P_0(t)$.
\item[iii] $K(n,1)=1$ for all $n$. Then the PMF reduces to a geometric distribution $(1-q)q^n$ with $q=\beta(t)$  
\item[iv] $K(n,2)=n+1$ for all $n$;
\item[v] If $r<1$, then $K(n,r)<1$ for all $n\geq 0$. From Property ii and (\ref{PMF-I(t)}), we find that for each $t$, the PMF $P_n(t)$ is a monotone decreasing function of $n$, i.e., $P_n(t)\propto \beta(t)^n$, where $\beta(t)$ approaches 1 from below as $t$ becomes large.
\item[vi] If $1<r<2$, then $K(n,r)$ is a monotone increasing concave (i.e., convex cap) function of $n$, bounded by $1<K(n,r)<n+1$.
\item[vii] If $r>2$, then $K(n, r)$ is a monotone increasing convex function of $n$, bounded from below $K(n, r)>n+1$.
\item[viii] By using Stirling's approximation for factorials and the Gamma function,
      \begin{align}
      \Gamma(n+1)=n!\approx \sqrt{2\pi n}\left(\frac{n}{e}\right)^n,~~\mbox{for}~~n>1,
      \end{align}
      we can approximate $K(n, r)$, for $n>1$ and $r>1$, by
      \begin{align}
      K(n,r)\approx \sqrt{\frac{1}{2\pi}}\frac{(n+r-1)^{n+r-1}}{(r-1)^{r-1}{n^n}}.
      \end{align}
\end{itemize}
       
  Thus, if $r$ is small, e.g., $r\approx 1$, the distribution of (\ref{PMF-I(t)}), for any given $t$, slowly decreases towards  zero as $n$ increases.  Thus, the distribution has a \emph{long tail} (see Figures 7-12 of Part I).  Consequently, different \emph{sample paths} of $I(t)$ are expected to exhibit enormous disparities, as will be shown in the next section, where 
we present simulation results, which demonstrate huge spreads across different sample paths.

It goes without saying that the most decisive factor that shapes the epidemic pattern is the exponential parameter $a=\lambda-\mu$. However, for a given $a$, different simulation runs show enormous differences in their infectious patterns. This wild random behavior is perhaps well beyond what most epidemiologists are cognizant of.  In other words, we should accept that we won't be able to find out all factors that can explain why some cities or countries are experiencing more hardship than others in terms of the number of infections and  death tolls.  

This also implies that Japan and other countries that have observed a relatively small number of infections and casualties should be aware that they may not be as fortunate in new waves of the pandemic.  They should be well prepared for the worst case scenario in order to protect their citizens. 

In order to be able to provide specifically what will be the worst possible scenario, we need to come up with an accurate estimates of the model parameters $\lambda, \nu$ and $\nu$, by carefully analyzing real and reliable data, and make the most likely estimates (or ranges of estimate), which will be the main focus in Part IV \cite{kobayashi:2021c} of our report. 

\section{Simulation of Our Proposed Stochastic Model}
Simulations are often used when analytic techniques to estimate or predict the performance of a complex system are hard    
 to come by. So-called \emph{Monte-Carlo simulation} techniques, such as \emph{variance reduction techniques}, are often adopted to estimate a given performance measure accurately and efficiently. The main purpose of our simulation experiments here, however, is different from these situations in that we fortunately have obtained an exact analytic result of the system, i.e., we have found closed-form expressions for the PMFs $P_n(t)$'s by solving a partial differential equation.
 
We present here the results of various simulation runs primarily to better explain and demonstrate the validity and utility of our stochastic model. Plots of various sample paths of our simulation experiments should serve as a convincing and understandable evidence to support our model.
 
\subsection{Description of the simulation model}

We shall briefly describe how our simulator is designed for the benefit of some readers who may not be sufficiently familiar with probabilistic simulation \footnote{As for detailed discussion of simulation techniques, see e.g., \cite{kobayashi:1978}, Chapter 4, \cite{kobayashi-mark:2008}, Chapter 16.} of a Markov process such as the birth-death-immigration process.  We adopt the \emph{time-asynchronous} approach as opposed to the time-synchronous approach, which may be more suited to simulating events that occur periodically at predetermined moments, such as seen in time-synchronous communication systems. The time-asynchronous approach is also referred to as the \emph{event-scheduling} approach.  

In our model, there are three types of events; (A) an arrival of an infected person from the outside; (B) an infection caused by an infectious person; and (R) a recovery/removal/death of an infected person.  

The random process $I(t)$ defined by the differential equations (\ref{Diff_eqn_P_0(t)}) and (\ref{Diff_eqn_P_n(t)}) is a Markov process.  At an arbitrarily given time $t$, when $I(t)=n$, the time interval until the next event is a random variable (RV) $X$ with a negative exponential distribution of mean $1/\gamma(n)$, \footnote{Note that $[n(\lambda+\mu)+\nu]$ is the coefficient of the $P_n(t)$ term in the RHS of differential equation of $P_n(t)$  (\ref{Diff_eqn_P_n(t)}).} where
\begin{align}
\gamma(n)=n(\lambda+\mu)+\nu. \label{gamma(n)}
\end{align}
That is, the distribution function of the RV $X$ is
\begin{align}
F_X(x)\triangleq \mbox{Pr}[X\leq x]=1-e^{-\gamma(n) x}.  \label{RV-X}
\end{align}
Generation of instances of the random variable $X$ can be done by calling the \emph{random number generator} (RNG) that is available in almost all programming languages, including MATLAB.\footnote{The random number generator used today in MATLAB, Python, Pokemon, and others is what is known as Mersenne Twisted GFSR (general feedback shift register) sequence, developed by Makoto Matsumoto and Takuji Nishimoto (see ``Mersenne Twister: a 623-dimensionally equidistributed uniform pseudo-random number generator," \emph{ACM Transactions on Modelling and Computer Simulation}, January 1998). This algorithm generates a pseudo-random sequence of the period $2^{19937}-1$, and the implementation is abbreviated as MT19937. It uses a Mersenne prime number $M$, of the form $M=2^n-1$, for some integer $n$. Sometimes $n$ is restricted to a prime number, as well.}  
It will generate an instances $u$ of the random variable $U$, which is uniformly distributed between 0 and 1. Then we transform $u$ by the function $F_X^{-1}(\cdot)$, i.e., find $x$ such that $u=F_X(x)=1-e^{-\gamma(n) x}$. In other words, $x=\log (1-u)/\gamma(n)$. Since $u$ is uniformly distributed between 0 and 1, so is $1-u$, thus $x'\triangleq\log u/\gamma(n)$ can be used instead of $x$,

At the occurrence of an event, we classify it as one of the three types of events as follows: choose type-(A) with probability $P_A\triangleq \nu/\gamma(n)$; choose type-(B) with probability $P_B=n\lambda/\gamma(n)$; and choose type-(R) with probability $P_R=1-P_A-P_B=n\mu/\gamma(n)$.  Note that if $n=0$, then $P_A=1$, and $P_B=P_R-0$.

The rationale for this extremely simple event scheduling approach is that the Poisson process possesses the following beautiful properties: (a) the \emph{memoryless property}, (b) the \emph{reproductive additivity}, and (c) the \emph{decomposition} property.\footnote{The property (a) has to do with the memoryless property of the variable $X$ which has an exponential distribution. Assume that $Y$ time units have elapsed since the last arrival.  The time until the next arrival $R\triangleq X-Y$ has the same exponential distribution as $X$, regardless of $Y$. The property (b) means that when $m$ independent Poisson processes with rate $\lambda_k, k=1,2, \cdots, m$ are merged, the resulting stream is another Poisson process with rate $\sum_{k=1}^m \lambda_k$. The property (c) is an opposite of the property (b).  If a Poisson process with rate $\lambda$ is split into $m$ sub-streams, by assigning each arrival independently into the $k$th sub-stream with probability $p_k$, where $\sum_{k=1}^m p_k=1$.  Then the sub-streams are independent Poisson processes with rate $p_k\lambda$'s. For proofs of these properties, see e.g. \cite{kobayashi-mark-turin:2012}, pp. 403-405). } 

Thus, the core of our simulation program is as follows.
\begin{enumerate}
\item If $I(t)=0$, then $\gamma(0)=\nu$, hence, the only possible event to consider is a type-(A) event. The time until the next arrival, $x$, is determined by $x=\log u/\nu$, where $u\in [0, 1]$ is an output drawn from the RNG. Advance the clock time to $t'\triangleq t+x$, and increment $A(t)$ by 1 for $t\geq t'$.

\item If $I(t)=n\geq 1$, the next event can be any of the three types.  The time-interval until the next event is determined by $x=\log u/\gamma(n)$T, where $u\in[0,1]$ as defined above. Take another RNG output $u'\in[0, 1]$.  Classify the event as type-(A), if $0\leq u'\leq P_A$: classify it as a type-(B) event, if $P_A< u'\leq P_A+P_B$; and classify it as type-(R), if $P_A+P_B<u'\leq 1$  Advance the simulation clock to $t'\triangleq t+x$, 
and increment the counter $A(t), B(t)$ or $R(t)$ by one depending on the classified result is type (A), (B) or (R).

\item  Go back to step 1 or 2, depending $I(t')=0$ or $I(t')\geq 1$, and use $t'$ as a new clock time, and repeat the above steps.
\end{enumerate} 

\subsection{Percentiles plots for the expected spread of simulation curves}

\begin{figure}[thb]
 \begin{minipage}[t]{0.45\textwidth}
  \centering
  \includegraphics[width=\textwidth]{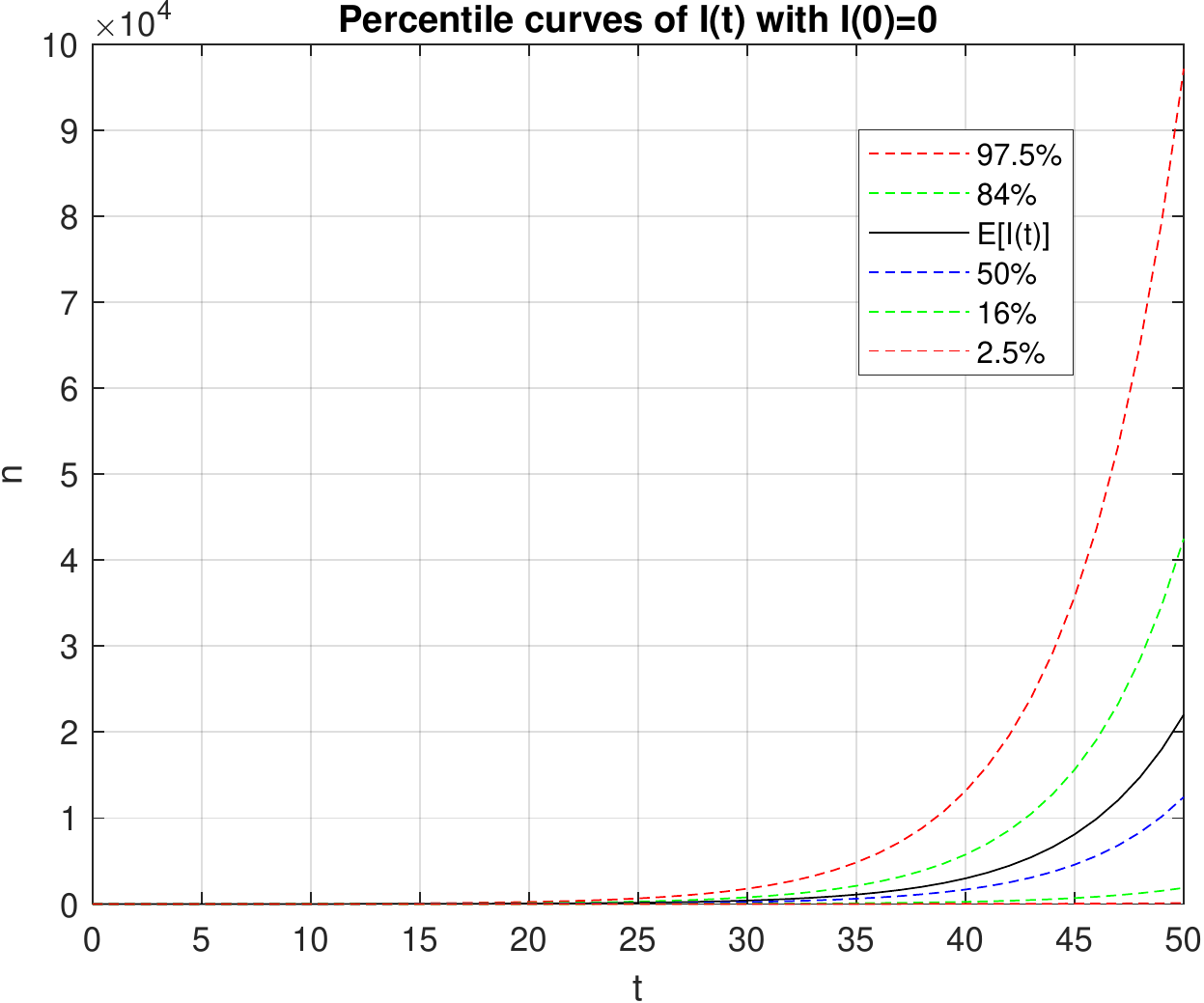}
  \caption{\sf\small Percentile curves of $P_n(t)$ of $I(t)$for $0\leq t \leq 50$, $\lambda=0.3, \mu=0.1, \nu=0.2.$}
  \label{fig:Percentile_curves_I(0)=0_t=50}
  \end{minipage}
  \qquad 
  \begin{minipage}[t]{0.45\textwidth}
  \centering
  \includegraphics[width=\textwidth]{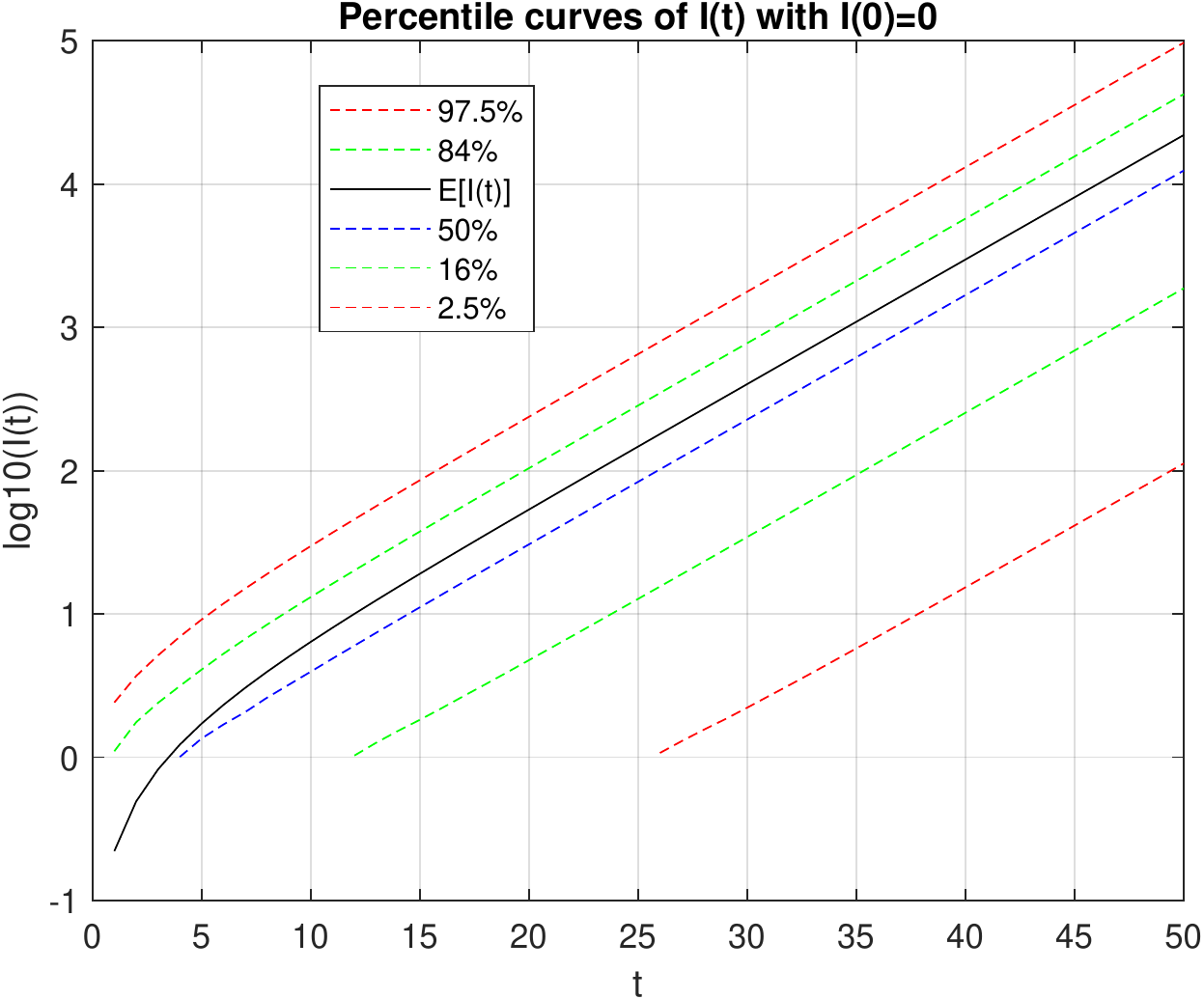}
  \caption{\sf\small Semi-log plot of the percentile curves for $0\leq t \leq 50$, $\lambda=0.3, \mu=0.1, \nu=0.2.$} 
  \label{fig:Semilog_Percentile_curves_I(0)=0_t=50}
  \end{minipage} 
\end{figure}

As we discussed in Part I, the ratio of the standard deviation of a given probability distribution to its mean, called the \emph{coefficient of variation} (CV), sometimes serve as a simple  and good measure of the dispersion of the distribution.  We obtained the CV of the distribution of $I(t)$ in the BDI model, as given in (74) in Part I.  

If the distribution of a random variable $X$ is close to a normal (or Gaussian) distribution, it is well known that realizations of $X$ fall within the mean $\pm\sigma_X$ with the probability 68.2\%. Similarly, the mean $\pm 2\sigma_X$ provides a 95.4\% confidence level, and the mean$\pm 3\sigma_X$ gives 98\% confidence level. But this rule of thumb does not apply to the NB($r,q$) with small $r$, because the distribution is highly skewed, far from being symmetrical around the mean.  A more accurate and reliable way is to compare simulation plots with the percentile curves of the NB($r,q$).  To that end, we calculate the \emph{cumulative distribution function} (CDF) by
\begin{align}
F_{\scriptscriptstyle BDI}(x, t)\triangleq \mbox{Pr}[N(t)\leq x]=\sum_{n=0}^{\left\lfloor{x}\right\rfloor} P_n(t), \label{CDF-I(t)}
\end{align}
where $P_n(t)$ is the PMF of (\ref{PMF-I(t)}) and $\left\lfloor{x}\right\rfloor$ is the largest integer, not exceeding $x$. Figure \ref{fig:Percentile_curves_I(0)=0_t=50} shows the curves where the CDF take 0.975, and  0.025 (in red dash curves), 0.84 and 0.16 (in green dash) and 0.5 (in blue dash).  The stochastic mean $E[I(t)]\triangleq\oI(t)$ is shown in black solid curve.

We can expect that if we conduct many simulation experiments, about 68\% of the sample paths will fall within the region between the green dashed curves, and about 95\% of the sample paths should fall within the range given by the two red dashed curves.  Roughly one half of the simulation curves should be above the blue dashed curve and the other half should be below this curve.  Note that the stochastic mean curve is appreciably above the median (50\%) curve: nearly by a factor of two.

Note that the top curve (the upper half of the 95\% confidence interval) climbs up to as large as 100,000 by the 50th day, while it is only 1,800 on the 30th day.  This is the power of the exponential growth, with which we are all familiar now by observing how rapidly the COVID-19's infections have grown in many parts of the world.

\subsection{Simulation results}

\begin{figure}[thb]
  \begin{minipage}[t]{0.45\linewidth}
  \centering
  \includegraphics[width=\textwidth]{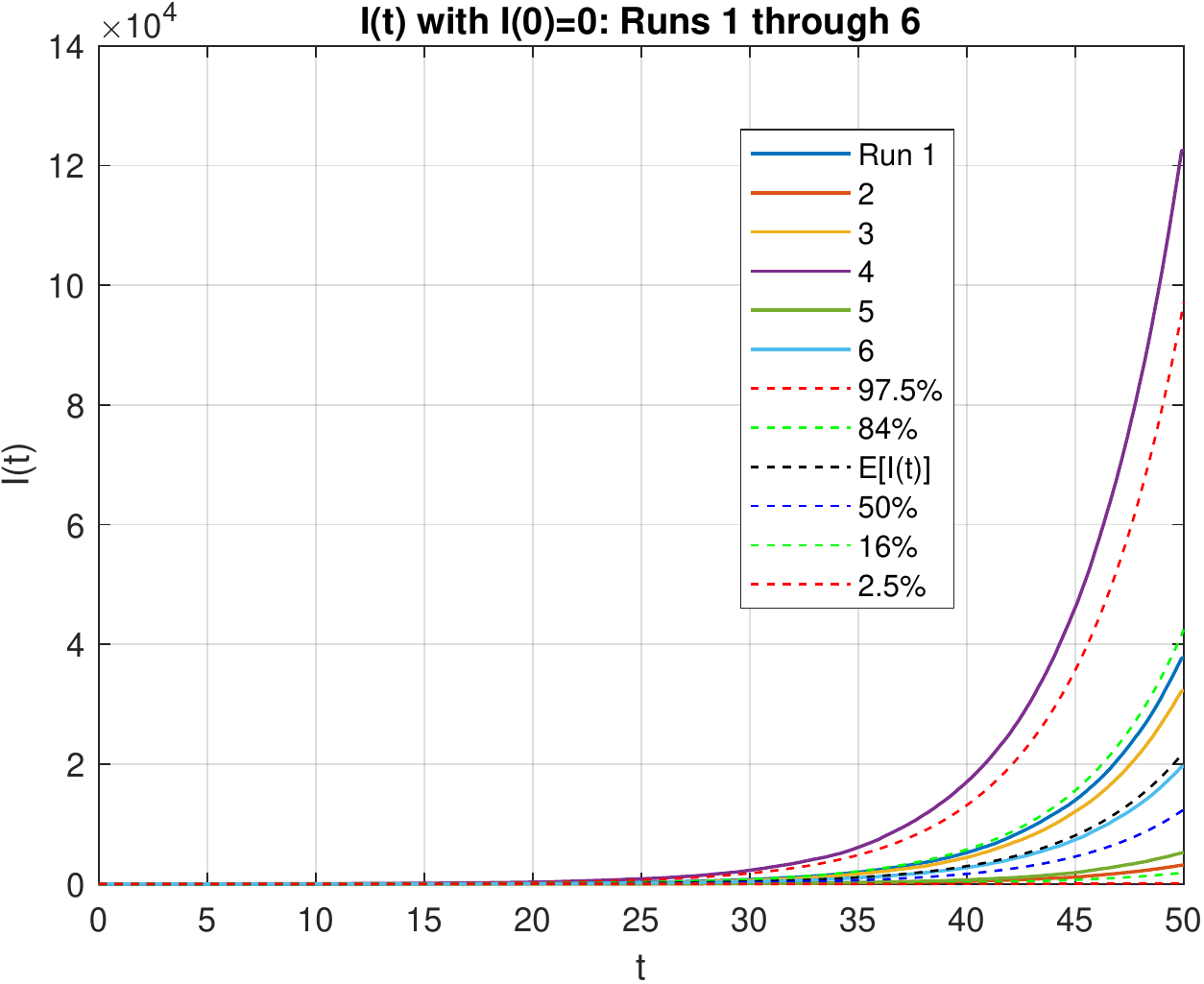}
  \caption{\sf\small Simulated $I(t)$ process, Runs 1-6; $\lambda=0.3, \mu=0.1, \nu=0.2.$}
  \label{fig:I(t)_Runs_1-6}
  \end{minipage}
  \qquad 
  \begin{minipage}[t]{0.45\linewidth}
  \centering
  \includegraphics[width=\textwidth]{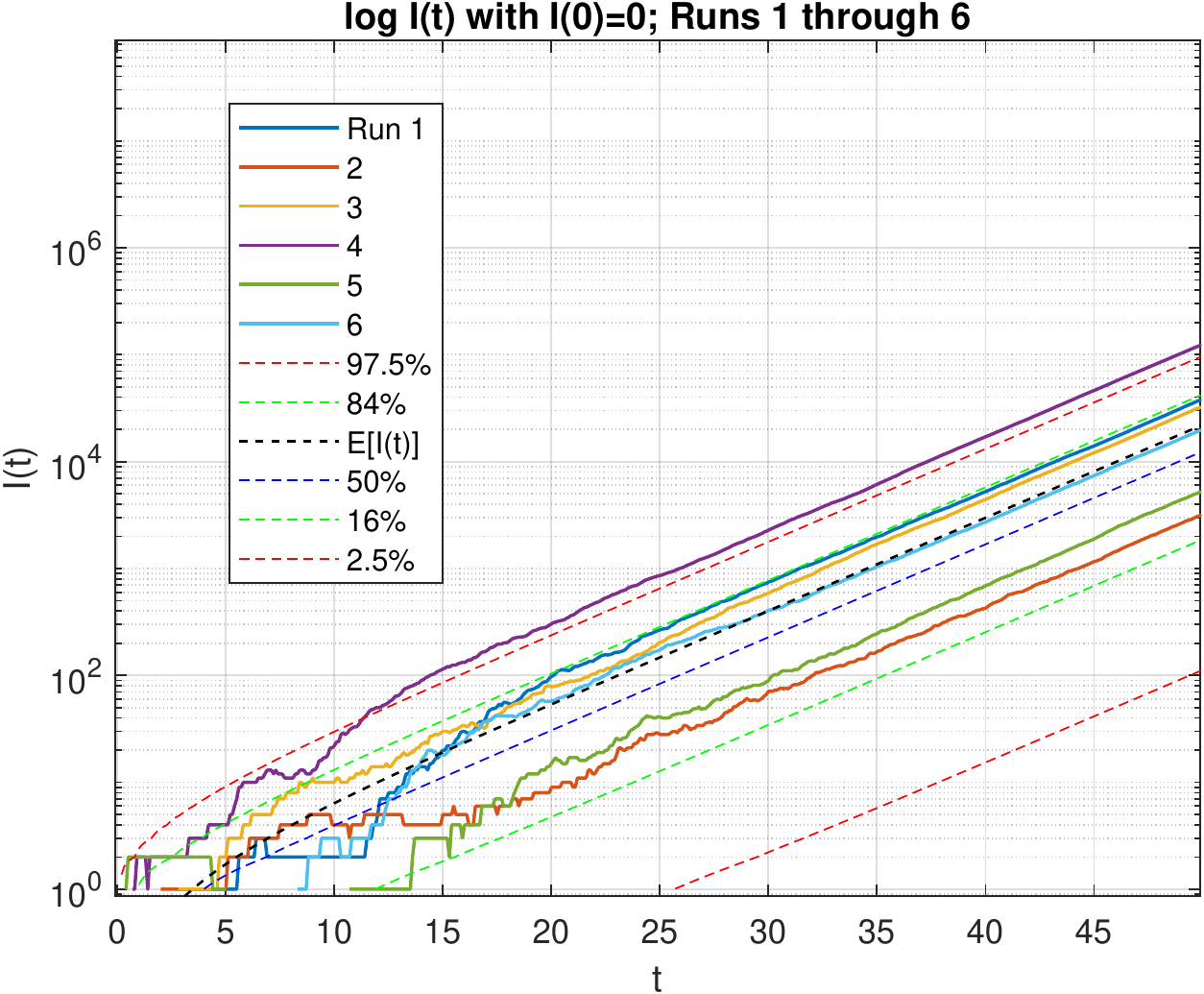}
  \caption{\sf\small Semi-log plots of simulated $I(t)$ process, Runs 1-6; $\lambda=0.3, \mu=0.1, \nu=0.2.$}
  \label{fig:Log_I(t)_Runs_1-6}
  \end{minipage}
  \hfill
  \begin{minipage}[b]{0.45\linewidth}
  \centering
  \includegraphics[width=\textwidth]{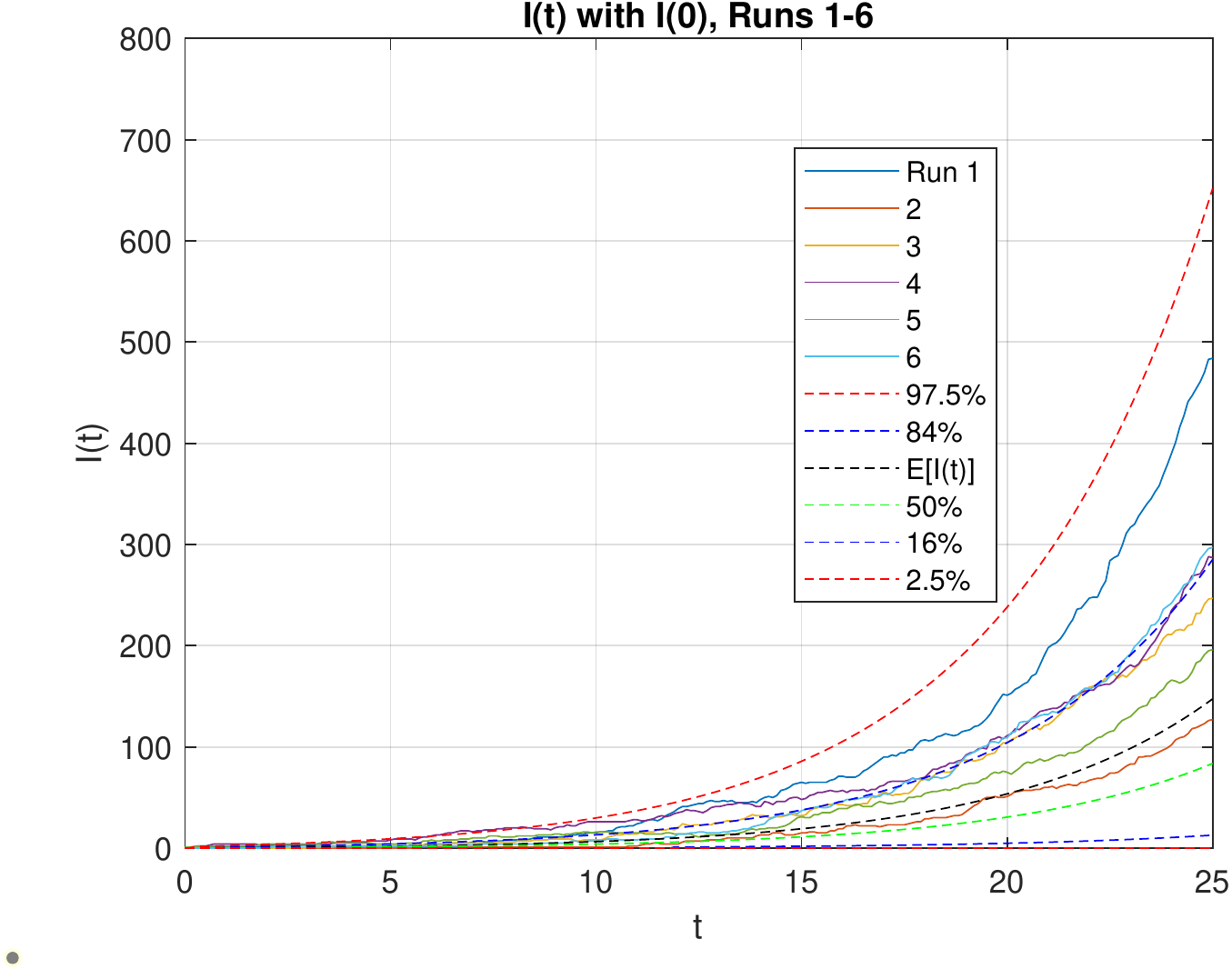}
  \caption{\sf\small Plots of simulated $I(t)$ process in the first 25 days, Runs 1-6; $\lambda=0.3, \mu=0.1, \nu=0.2.$}
  \label{fig:I(t)_Runs_1-6_t_25}
  \end{minipage}
  \qquad 
  \begin{minipage}[b]{0.45\linewidth}
  \centering
  \includegraphics[width=\textwidth]{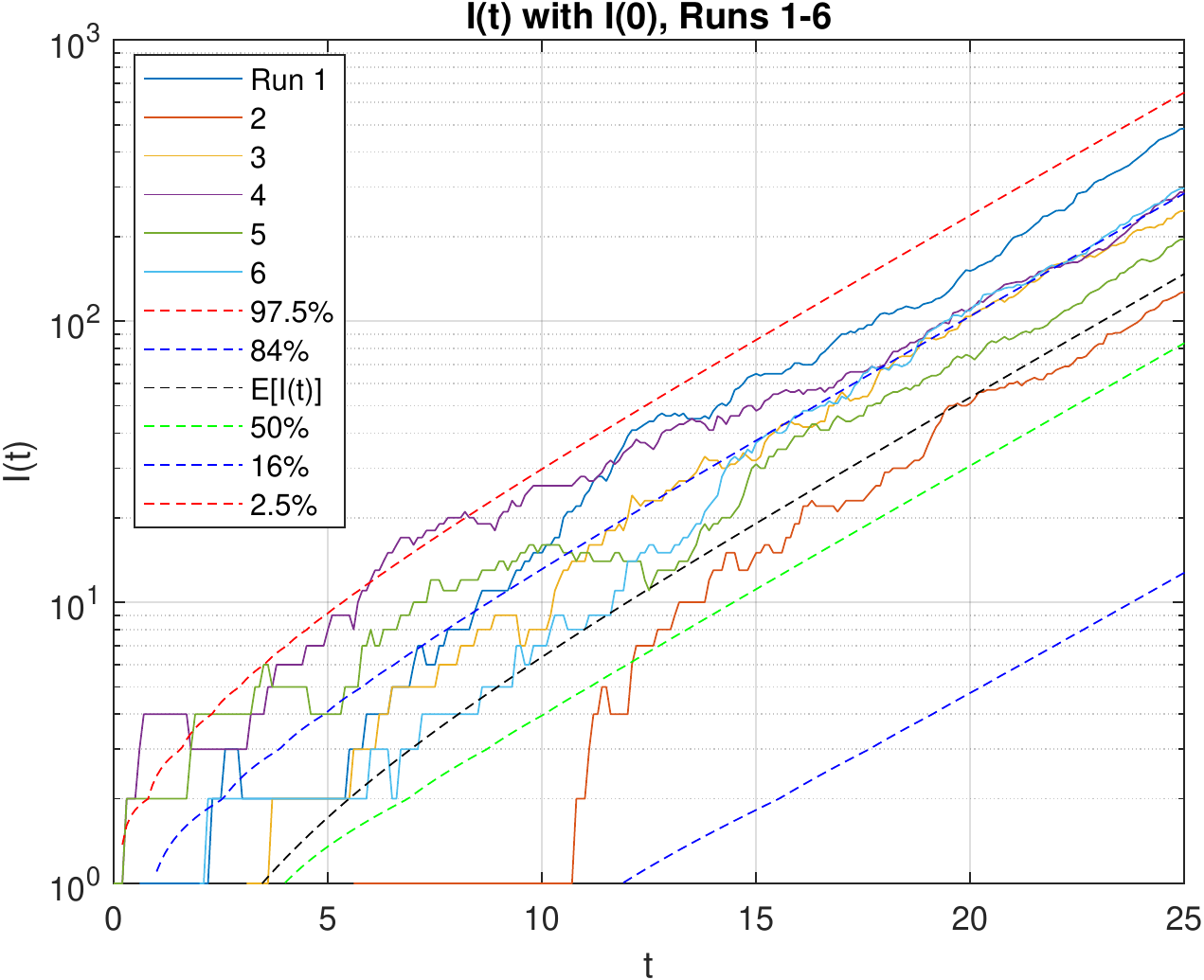}
  \caption{\sf\small Semi-log plots of the $I(t)$ process, Runs 1-6; $\lambda=0.3, \mu=0.1, \nu=0.2.$}
  \label{fig:Log_I(t)_Runs_1-6_t_25}
  \end{minipage}
\end{figure}

\subsubsection{Simulation of the process $I(t)$, the number of infections at time $t$}

We present the results of six consecutive simulation runs (in solid curves) in Figure \ref{fig:I(t)_Runs_1-6}, where we superimpose the percentile curves (in dashed curves)
obtained in the previous section. The expected value $\oI(t)=\Ex[I(t)]$ of (\ref{mean-I(t)}) is shown in a black dashed curve that ends 
at $t=500$ with the value $\frac{0.2}{0.2} \left(e^{0.2*500}-1\right)=e^{10}-1\approx 2.20\times 10^4$. Three runs (Runs 4, 1 and 3) are clearly above $\oI(t)$, 
and three runs (Runs 6, 5 and 2) lie below $\oI(t)$, although Run 6 (in light blue) is very close to $\oI(t)$. This may not necessarily be a typical situation, because one half of the simulation runs, on average, should be above the ``median curve", i.e., 50 percentile curve (shown in blue dash). Five of the six runs are within the 68\% confidence interval, (between the two green dashed curves).  

The most important observation to make here is the enormous disparity among the six simulations. Run 4 (in cyan) has as many as $1.22\times 10^5$ infected people at $t=500$ [days], whereas Run 2 (in red) has only $\approx 3\times 10^3$. Their ratio, therefore, is more than a factor of 40.  If we run the simulator many times, for instance, the ratio of the numbers of the infected of the worst (e.g.,97\% level) vs. the luckiest  (e.g., 2.5\% level) cases can be as large as 1,000 (see \cite{kobayashi:2020s}, slide \#33.).  

If we plot the same simulation runs using the semi-log scale, as shown in Figures \ref{fig:Log_I(t)_Runs_1-6_t_25}, we can see more clearly their behaviors in the initial rising phase (i.e., for small $t$) of these simulation runs of the stochastic process $I(t)$.  From the plots in the semi-log scale, we can see that once $I(t)$ has reached the level of $\approx 100$, the infected number in each run grows in a more or less deterministic fashion with a slope of 0.087 \footnote{~~~ $\log_{10}e^{at}=\log_{10}e\times 0.2t=0.087 t$.}.  This is because once the $I(t)$ has reached $\approx 100$, then a ``(weak) law of large numbers" (As for detailed discussion on weak vs. strong law of large numbers, see e.g. \cite{kobayashi-mark-turin:2012}, pp.298-300.) will set in, because $I(t)$ is the sum of many statistically independent and identical infection processes, and exhibits a more stable and predictable behavior than in the initial phase.  In other words, the large variance among different sample paths of the BDI process is due to more erratic and unpredictable behavior in the early phase of the stochastic process $I(t)$.  In the initial phase the random arrivals of the infected from outside, as well as the fluctuations in the internal infection (a branching process) and the recovery/death, all contribute to the stochastic behavior of $I(t)$ because $I(t)$ is still small.  As $I(t)$ grows the effect of external arrivals will become negligible as long as $r=\frac{\nu}{\lambda}$ is small, i.e., the order of unity or less.

As remarked in the present author's keynote speech \cite{kobayashi:2020s}, it may be worth noting that the BDI process model can be also used to explain the enormous disparity we observe in the wealth among different individuals of similar income levels, similar expenditure, and similar intelligence and knowledge in investment, thus they have a more or less similar investment portfolio in their initial phase.  After 30 or 40 years after beginning their careers after schooling, some may become multi-millionaires (or even billionaires), whereas some end up with a life of modest means, if not living hand to mouth.  The present author believes that this type of disparity in wealth distribution can be explained by applying the BDI process model. \footnote{ In this analogy between the infection process and the individual's wealth growth, we interpret the infection rate $\lambda$ as the ROI (return on investment), and the recovery/death rate $\mu$ as the rate of expenditures. The rate of arrivals of the infected from outside $\nu$ can be translated into the amount of additional money that can be put into investment.  The parameter $a=\lambda-\mu$ is equivalent to the net growth rate of the wealth.  Given that all these parameters are almost identical among different individuals,  the disparity in their wealth growth depends, to a large extent, on their luckiness or unluckiness in the early phase of their investments, because the growth or decrease of wealth will be dictated by randomness in success or failure of investment, unexpected large expenditure or loss, and the availability of new fund for investment.  Once the wealth exceeds a certain level, for instance, a million dollars, the future growth is more or less predictable from their investment strategy and expenditure because failures in some investments will average out with successes in other investments.  Unavailability of new fund for investment will not affect much compared with the early phase when the wealth is small.  The net asset growth will behave more or less predictably as a deterministic model can tell.  To the best of the author's knowledge the applicability of the BDI process model to explain the disparity in wealth distribution seems a novel approach.}.
 
\begin{figure}
  \begin{minipage}[b]{0.45\linewidth}
  \centering
  \includegraphics[width=\textwidth]{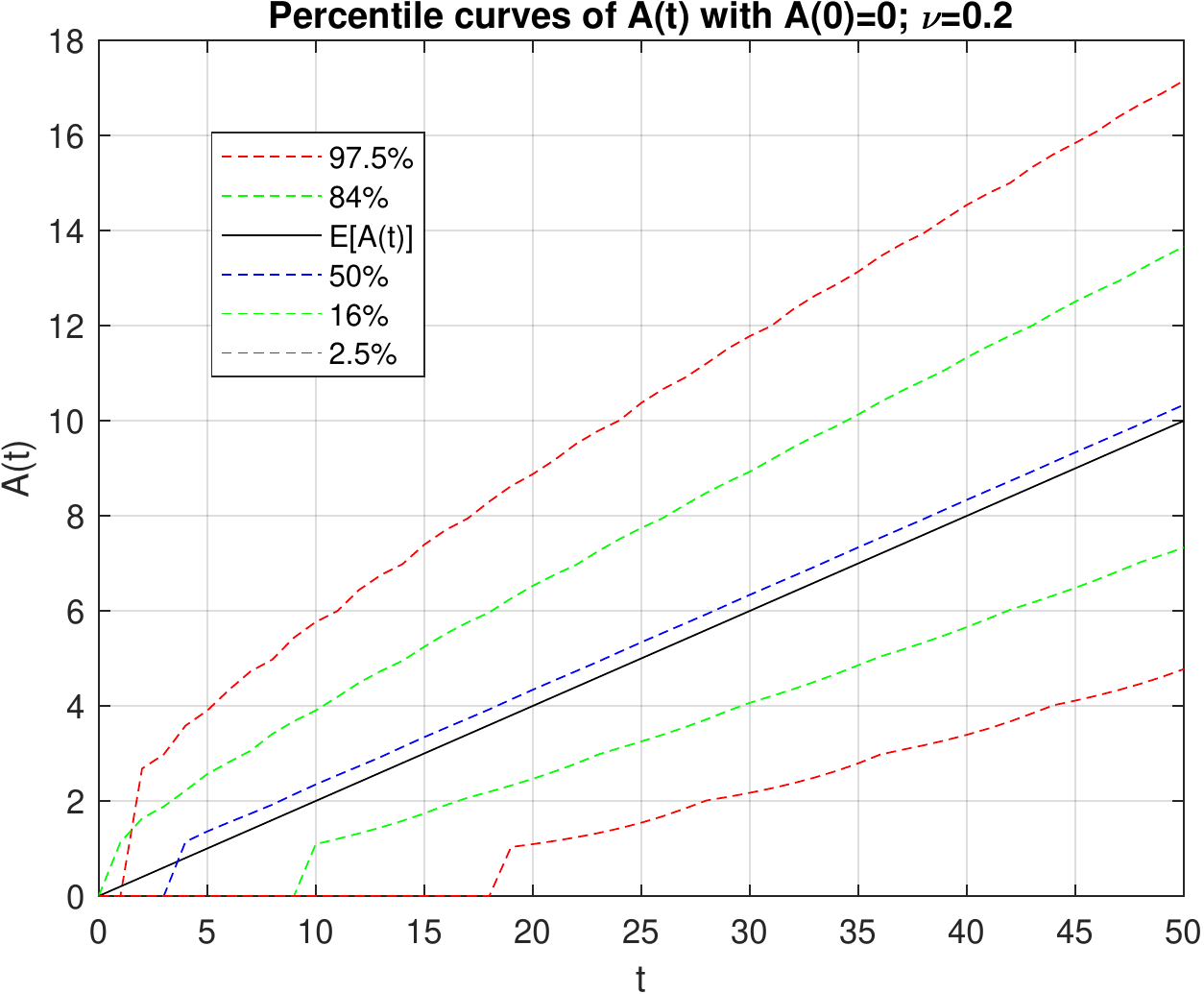}
  \caption{\sf\small Percentile curves of Poisson process $A(t)$ of rate $\nu=0.2.$}
  \label{fig:Percentiles_A(t)_nu=0.2}
  \end{minipage}
  \qquad
  \begin{minipage}[b]{0.45\linewidth}
  \centering
  \includegraphics[width=\textwidth]{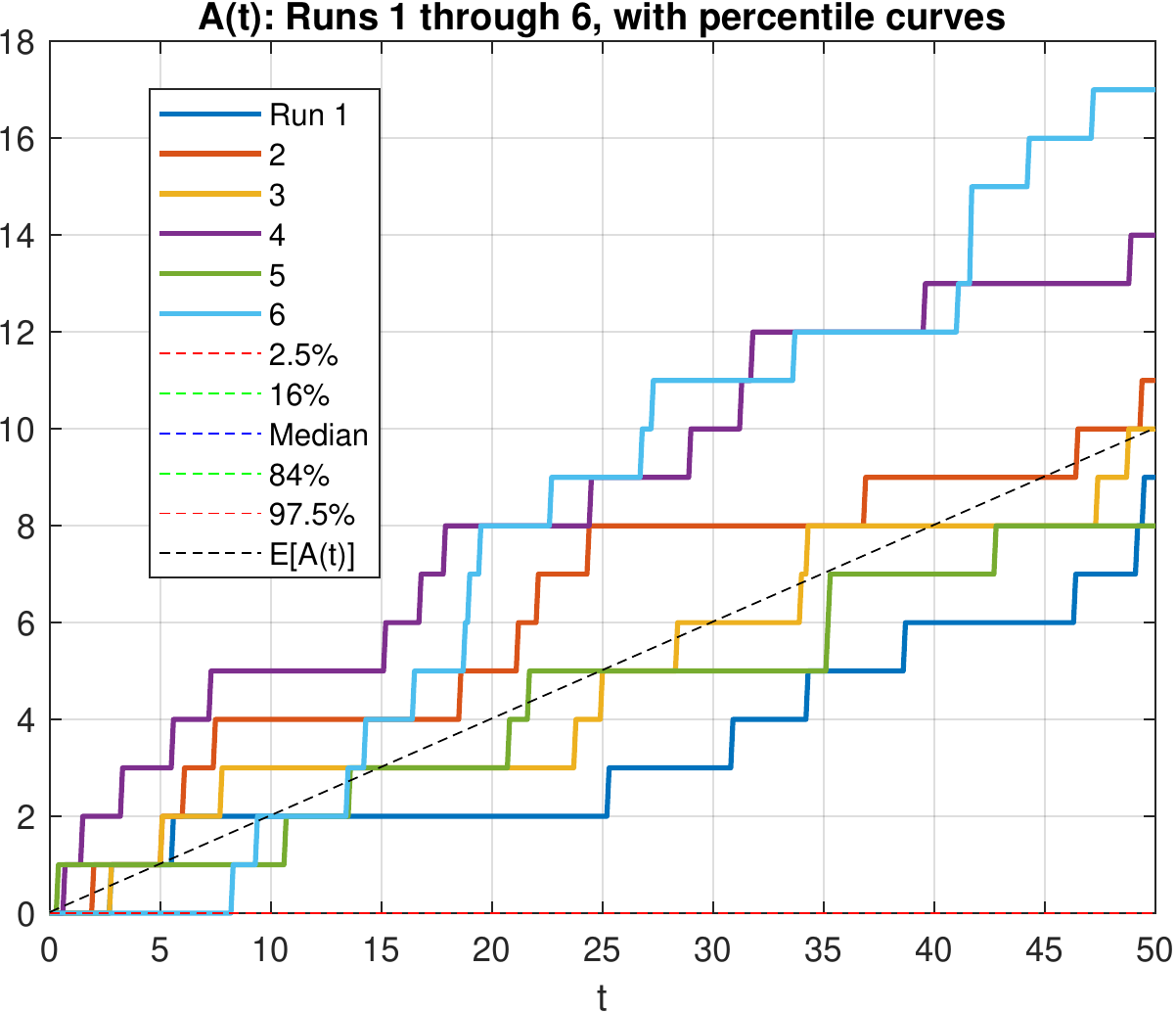}
  \caption{\sf\small External arrival process $A(t)$, Runs 1-6.}
  \label{fig:A(t)_Runs_1-6}
  \end{minipage}
\end{figure}

\subsubsection{Simulation of the external arrivals $A(t)$}

Figure \ref{fig:A(t)_Runs_1-6}  show the arrival patterns.
The Poisson distribution of mean $\lambda$ has the variance equal to the mean: $\sigma^2=\lambda$; and for large 
$\lambda$, its CDF (cumulative distribution function) converges to that of the normal (or Gaussian distribution);  Thus, 68 percent and 95 percent confidence levels could be well approximated by the $\lambda \pm \sqrt{\lambda}$, and the $\lambda\pm 2\sqrt{\lambda}$, respectively.  

The coefficient of variation of the Poisson process is given by
\begin{align}
c_{\scriptscriptstyle A(t)}=\frac{1}{\sqrt{\nu t}}, \label{CV_A(t)}
\end{align}
which converges to zero as $t\to\infty$.  Thus, the arrival process $A(t)$ behaves much more predictably than the process $I(t)$, whose coefficient of variation remains on the order of unity at all $t$.
\begin{align}
c_{\scriptscriptstyle I(t)}=\frac{1}{\sqrt{r\beta(t)}}, \label{CV_I(t)}
\end{align}
where $r=\frac{\nu}{\lambda}$ and $\beta(t)\leq 1$ is defined in (\ref{45-Part-I}). As  $t\to\infty$, $\beta(t)\to 1$, thus, the coefficient of deviation (CV) converges to $\sqrt{\frac{\lambda}{\nu}}$, which, in our running example is $\sqrt{0.3/0.2}\approx 1.225$ as we discussed in Part I, p. 19, Example 3.

\subsubsection{Simulation of the processes $B(t)$ and $R(t)$}

\begin{figure}[thb]
  \begin{minipage}[t]{0.45\linewidth}
  \centering
  \includegraphics[width=\textwidth]{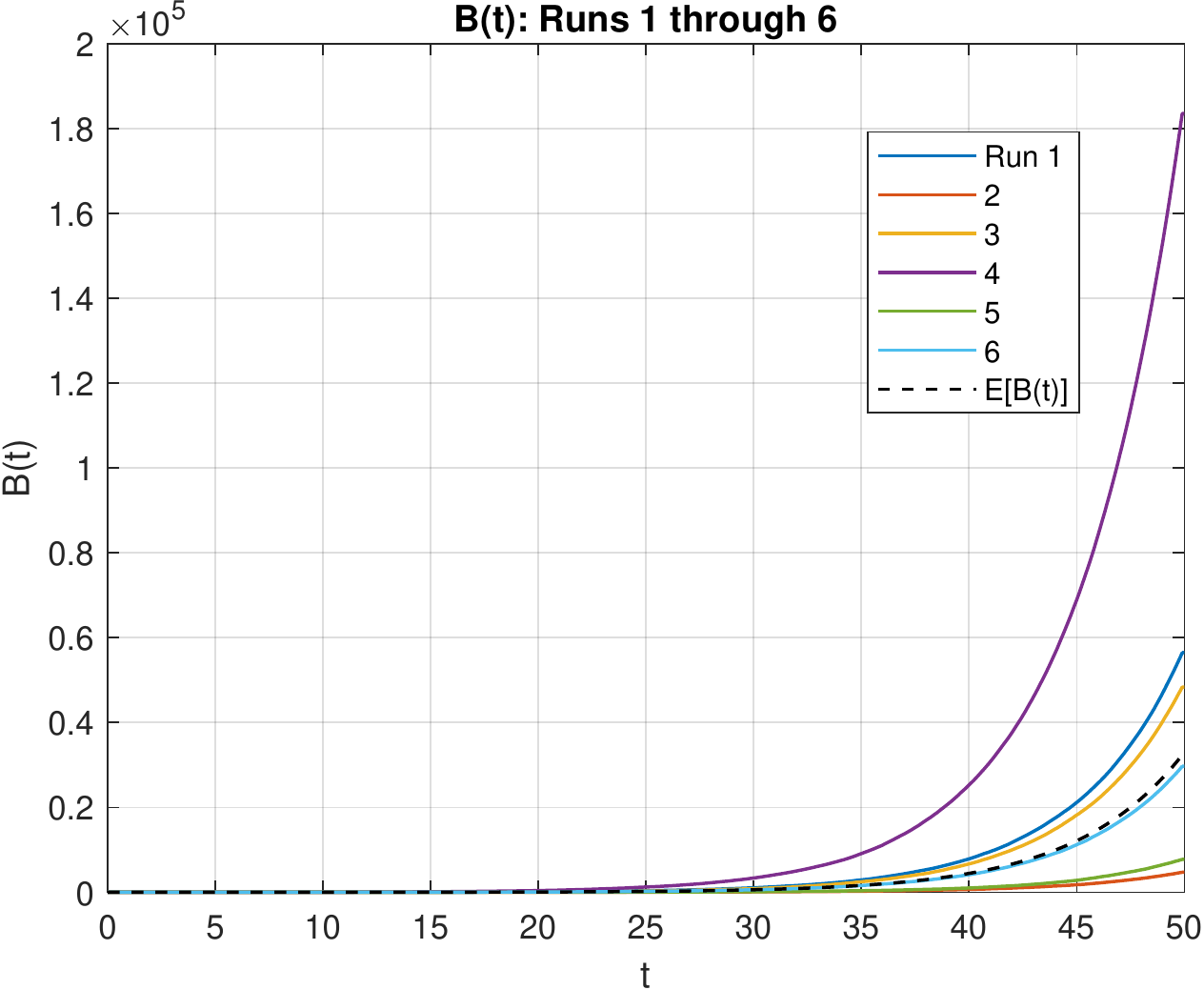}
  \caption{\sf\small The process $B(t)$: Runs 1-6.}
  \label{fig:B(t)_Runs_1-6}
  \end{minipage}
  \qquad 
  \begin{minipage}[t]{0.45\linewidth}
  \centering
  \includegraphics[width=\textwidth]{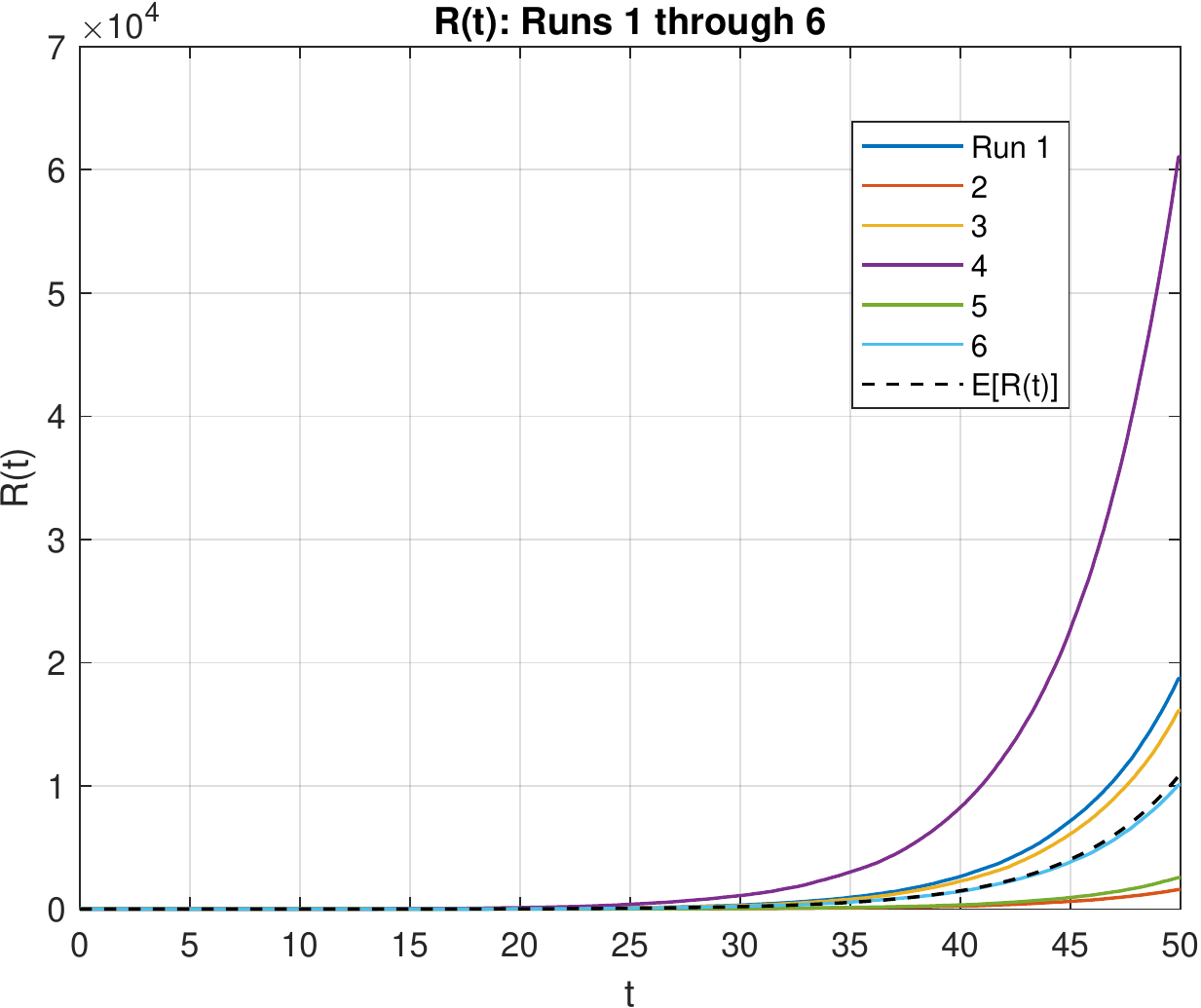}
  \caption{\sf\small The process $R(t)$: Runs 1-6.}
  \label{fig:R(t)_Runs_1-6}
  \end{minipage}
 \end{figure}
 
 Although we were able to obtain the time-dependent probability mass function (PMF) of $I(t)$, it seems rather difficult to obtain closed form expressions of PMFs for the processes $B(t)$ and $R(t)$, although we can find their PGFs.  It appears that we need to be content with approximate PMFs of $B(t)$ and $R(t)$.  The technique we will explore is the saddle-point integration method, which Bernhard Riemann (1826-1866) pioneered in his pursuit of the famous 1859 conjecture, known as the Riemann Hypothesis \footnote{See, e.g., Harold M. Edwards, \emph{Riemann's Zeta Function}, Dover Publications, Inc, 1974. Also \url{http://hp.hisashikobayashi.com/towards-a-proof-of-the-riemann-hypothesis-rh/} and references therein.}. We will report our full discussion of the processes $B(t)$ and $R(t)$ in Part V \cite{kobayashi:2021d}.

In the present section we only show the results of the above six simulation runs, together with the stochastic means $\oB(t)$ and $\oR(t)$, which we obtained in Part I, viz:
\begin{align}
\oB(t)&=\frac{\lambda}{a}(\oI(t)-I_0-\oA(t), \label{E[B(t)]}\\
\oR(t)&=\frac{\nu}{a}(\oI(t)-I_0-\oA(t)), \label{E(R(t)]}
\end{align}

As we can see, the process $B(t)$ and $R(t)$ also exhibit enormous variations across the runs. They are positively correlated with each other and with the $I(t)$ process, as well. Run 4 (magenta), Run 1 (blue) and Run 3 (yellow) are well above the mean curve, and Run 5 (green) and Run 2 (red) are below the mean in all the three processes $B(t), R(t)$ and $I(t)$.  Run 6 (cyan) is just below the mean. The fact that these three processes behave in a similar fashion is quite expected, because both $\frac{dB(t)}{dt}$ and $\frac{dR(t)}{dt}$ are, on average, proportional\footnote{New internal infections (excluding the new arrivals from the outside), occur at the rate of $\lambda I(t)$, and recoveries/removals/deaths occur at the rate of $\mu I(t)$. Their expected values are related by the differential equations
\begin{align}
\frac{d\oB(t)}{dt}=\lambda\oI(t)~~ \mbox{and}~~\frac{d\oR(t)}{dt}=\mu\oI(t),
\end{align}
as shown in Part I, page 11 (27) and (31).} to $I(t)$ at a given time.  

Referring to (\ref{I(t)-def}), $A(t)$ is much smaller than $I(t), B(t)$ and $R(t)$, except for the initial period, thus we have an approximate formula
\begin{align}
I(t)\approx B(t)-R(t),
\end{align}

These large variations we observe in the processes $B(t)$, $R(t)$ and $I(t)$ are all consequences of the built-in \emph{positive feedback loop} inherent to the internal infection process $B(t)$, which is a \emph{branching process} \footnote{ See \url{https://en.wikipedia.org/wiki/Branching_process} and references therein.} and gives rise to exponential growth $\exp(at)$.

\subsection{Other important statistical data}
In this section we discuss two additional topics for the benefits of the readers.  One is how to derive the  number of new infections for each day; the second is how to estimate the number of deaths.

\subsubsection{Number of daily new infections}
\begin{figure}[thb]
  \begin{minipage}[t]{0.45\linewidth}
  \centering
  \includegraphics[width=\textwidth]{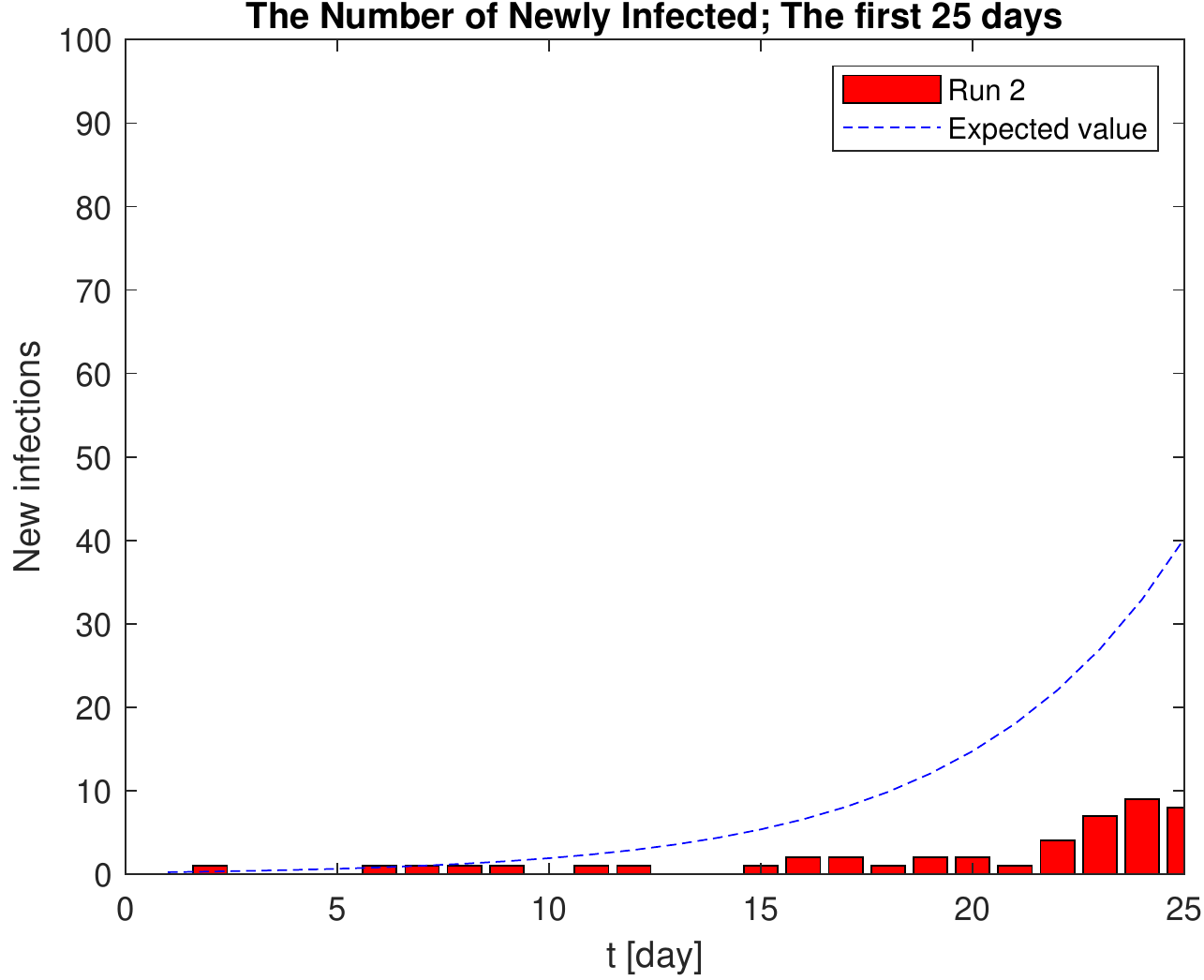}
  \caption{\sf\small New infections on Daily-basis, Run-2, $\leq t\leq 25$.}
  \label{fig:Daily_Run2_t=25}
  \end{minipage}
  \qquad 
  \begin{minipage}[t]{0.45\linewidth}
  \centering
  \includegraphics[width=\textwidth]{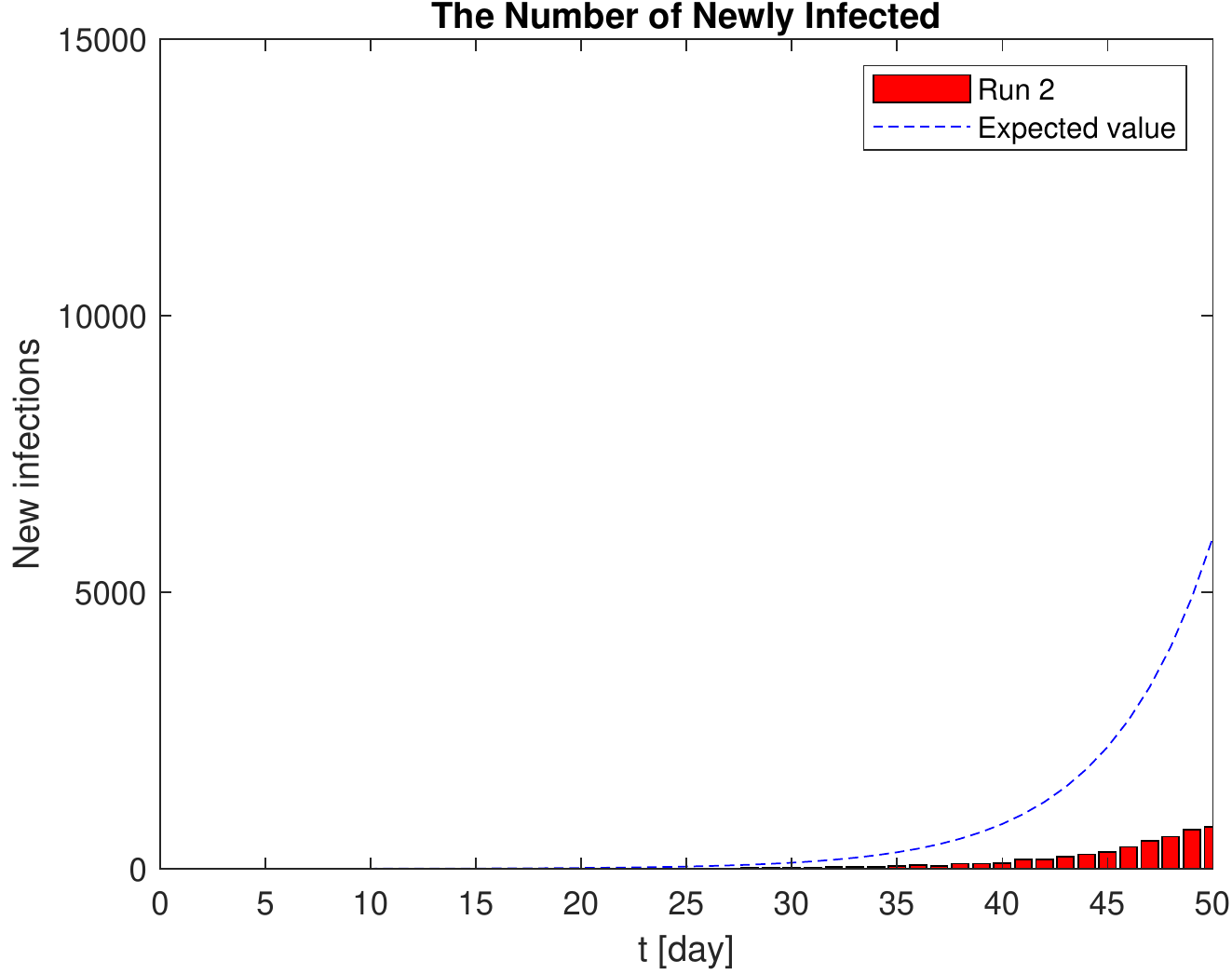}
  \caption{\sf\small New infections on Daily-basis, Run-2, $\leq t\leq 50$.}
  \label{fig:Daily_Run2_t=50}
  \end{minipage}
  \hfill
  \begin{minipage}[t]{0.45\linewidth}
  \centering
  \includegraphics[width=\textwidth]{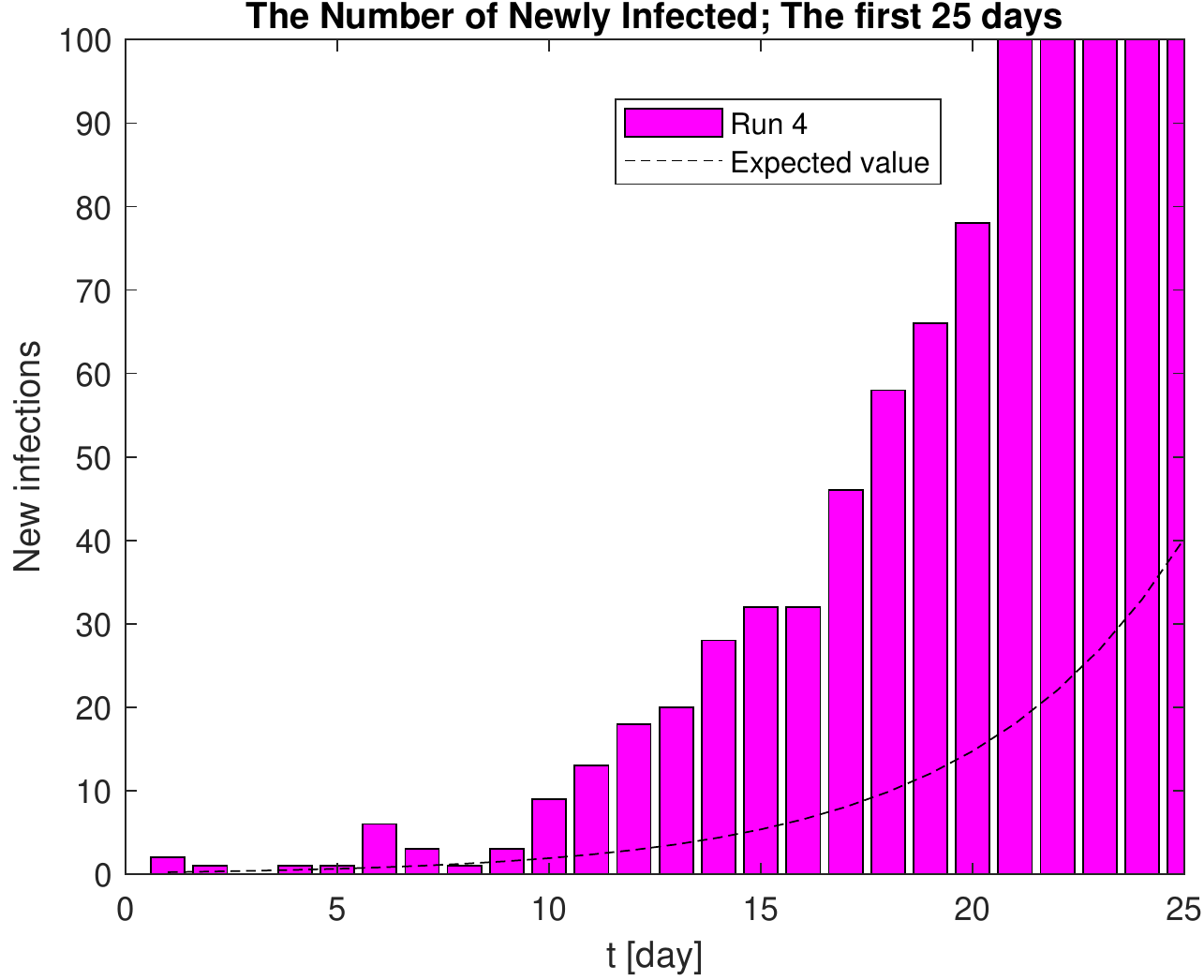}
  \caption{\sf\small New infections on Daily-basis, Run-4, $\leq t\leq 25$.}
  \label{fig:Daily_Run4_t=25}
  \end{minipage}
  \qquad 
  \begin{minipage}[t]{0.45\linewidth}
  \centering
  \includegraphics[width=\textwidth]{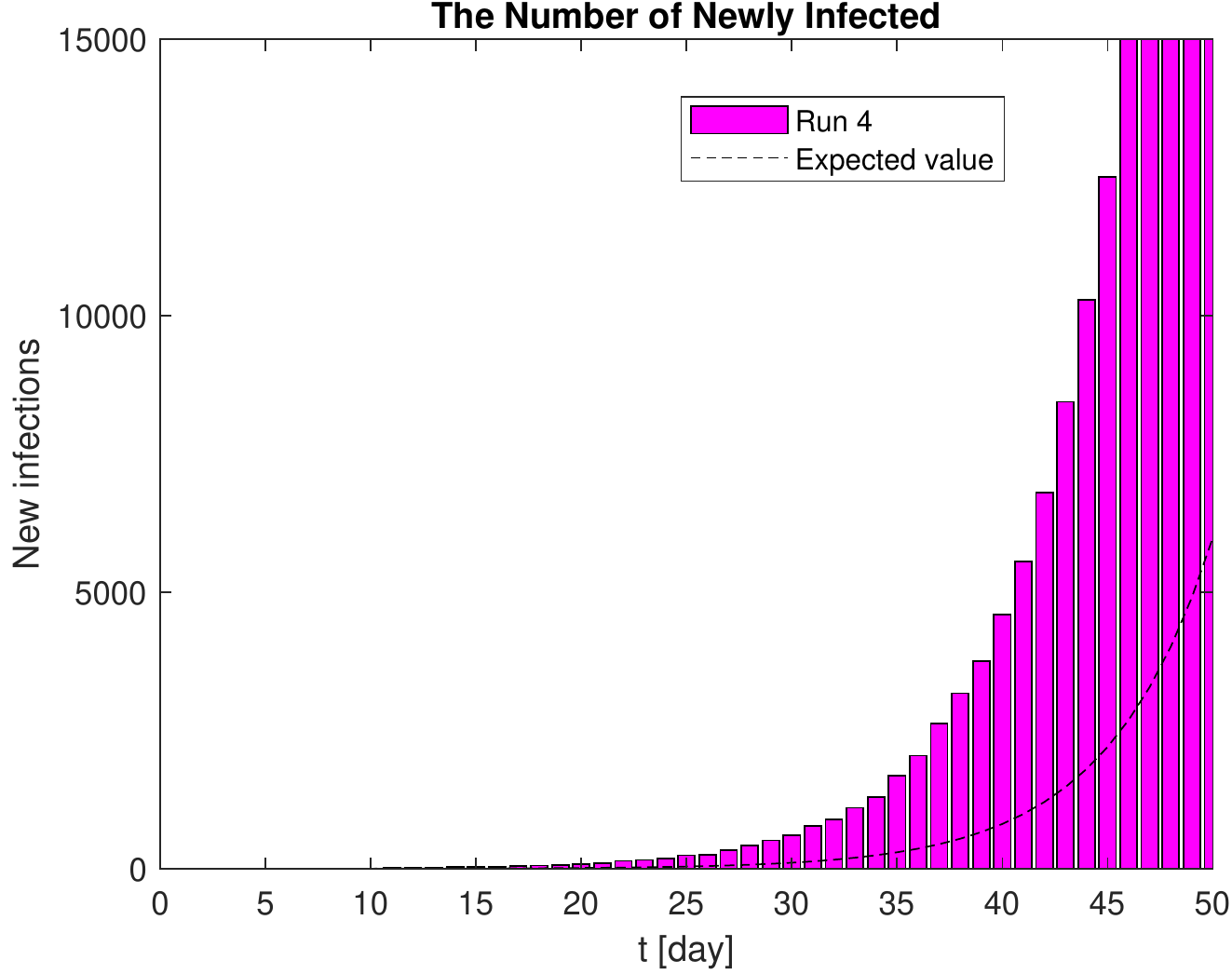}
  \caption{\sf\small New infections on Daily-basis, Run-4, $\leq t\leq 50$.}
  \label{fig:Daily_Run4_t=50}
  \end{minipage}
\end{figure}
\begin{figure}[bht]
  \begin{minipage}[b]{0.45\linewidth}
  \centering
  \includegraphics[width=\textwidth]{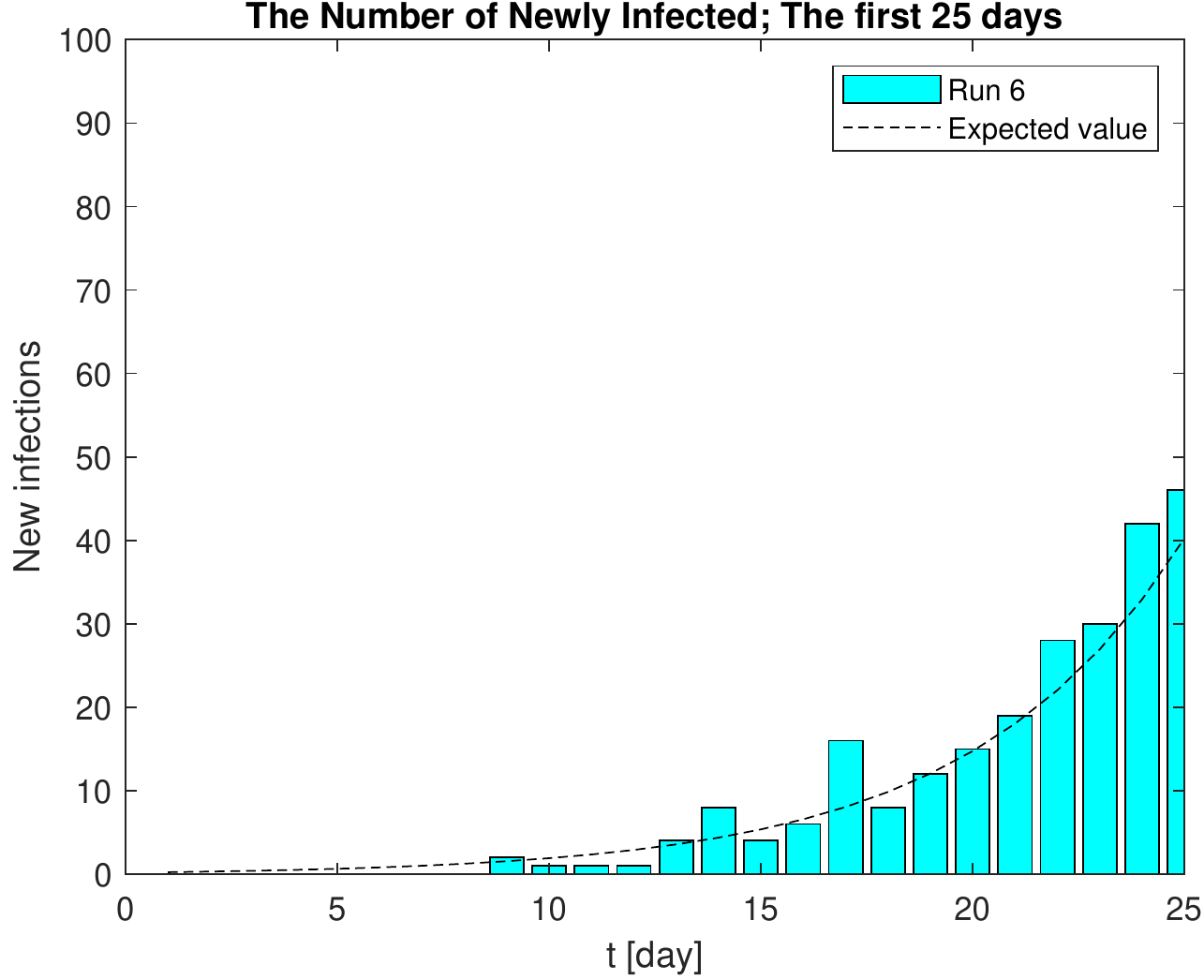}
  \caption{\sf\small New infections on Daily-basis, Run-4, $\leq t\leq 25$.}
  \label{fig:Daily_Run6_t=25}
  \end{minipage}
  \qquad 
  \begin{minipage}[b]{0.45\linewidth}
  \centering
  \includegraphics[width=\textwidth]{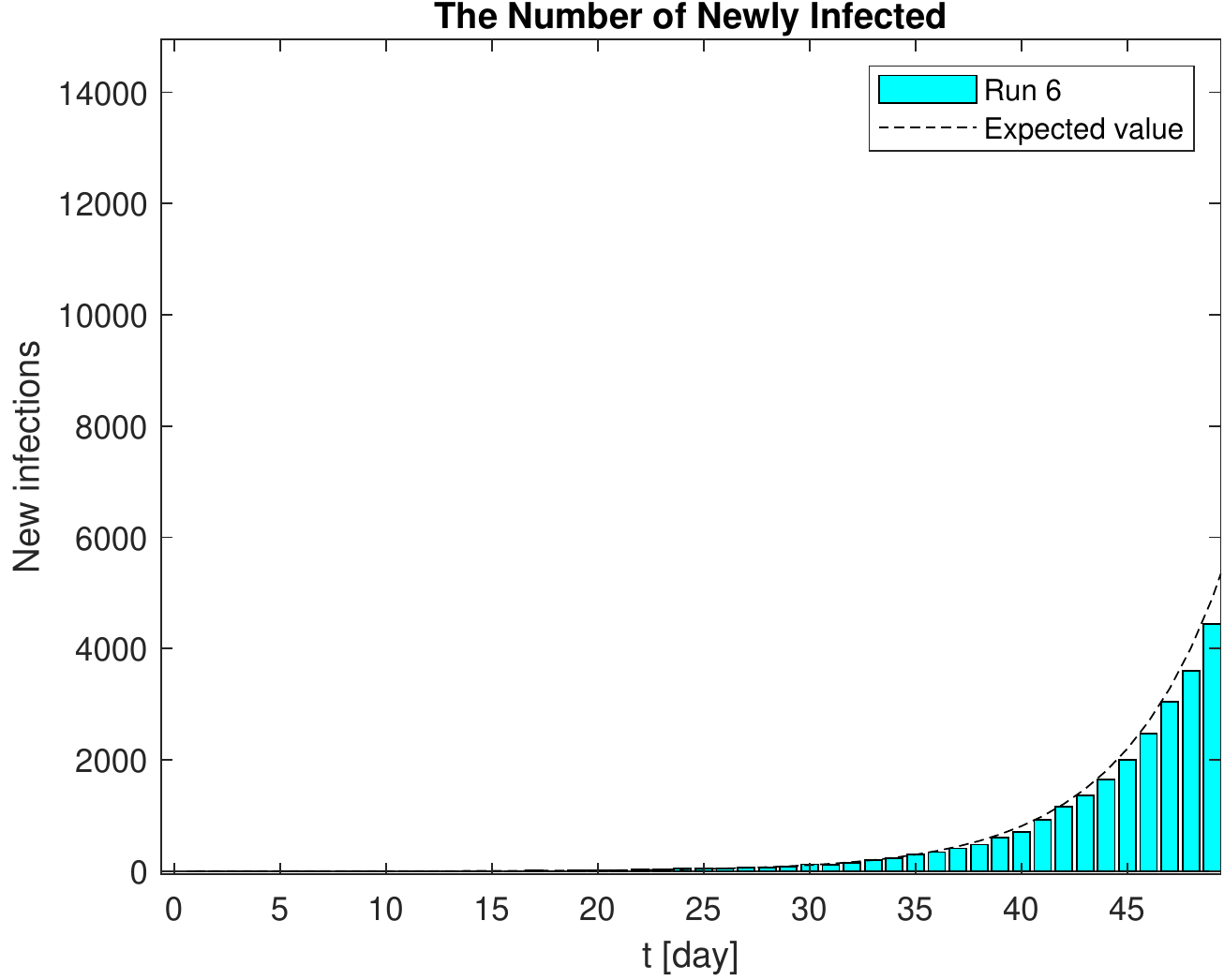}
  \caption{\sf\small New infections on Daily-basis, Run-4, $\leq t\leq 50$.}
  \label{fig:Daily_Run6_t=50}
  \end{minipage}
 \end{figure}
 
We take ``day"  as the time unit, as we do in our running examples throughout this report. We choose to interpret the  interval $t\in(0, 1)$ as the 1st day, then the \emph{number of newly infected persons} reported on the $t$th day, denoted as $N(t)$, should be computed as follows:
\begin{align}
I_{new}[t]=B(t)-B(t-1)+ A(t)-A(t-1),  \label{new-infections}
\end{align}
By substituting the $\oB(t)$ and $\oA(t)$ obtained in Part I, we readily find the stochastic mean or the expected value of $I_{new}[t]$:
\begin{align}
\oI_{new}[t]\triangleq \Ex[I_{new}[t]]=\frac{\lambda\nu}{a^2}(1-e^{-a})e^{at}-\frac{\lambda\nu}{a}.
\end{align}

For our running example of $a=\lambda-\mu=0.3-0.1=0.2$, and $\nu=0.2$. Figures \ref{fig:Daily_Run2_t=25} through \ref{fig:Daily_Run6_t=50} show the plots of the first 25 days and 50 days of three runs: Run-2, Run-4 and Run-6. These simulations correspond to shown in the previous figures concerning the processes $I(t), B(t), R(t)$ and use the same colors as used in the curves.  Run-2 (red) is the lowest; Run-4 (magenta) is the highest, exceeding the expected value $\Ex[I_{new}[t]]$ by factor of 4 or 5; Run-6 (light blue) is close to the expectation.  We observe enormous variations among the 6 runs. 
Since we have not yet obtained the PMFs for $B(t)$, we cannot compute the percentile curves at this point. But from these simulation runs, the variations among different runs seem even more pronounced than we observed in $I(t)$. 

Note that for the short range $1\leq t\leq 25$, the vertical axis range is [0, 100], whereas for $1\leq\ t \leq 50$, the range is expanded by a factor of 150.  This is consistent with the exponential growth of $\exp(a t)$.  From $t_1=25$ to $t_2=25$, it will grow by the factor of $\exp(a(t_2-t_1))=\exp(0.2\times 25)=e^5=148.4$

\subsubsection{Cumulative number of deaths}

The most important aspect of any model of an infectious disease should be how to estimate or predict the number of deaths. In the current COVID-19 epidemic, the case fatality rate is reportedly less than one percent for young people but will be much higher for 60 years or older, and those with comorbidity. In the present model, we assume a homogeneous population model, but the model can be generalized to multiple ``classes" (or types), by assigning the model parameters $\nu_c, \lambda_c$ and $\mu_c$ for different classes of infected population.
\begin{figure}[thb]
  \begin{minipage}[t]{0.45\linewidth}
  \centering
  \includegraphics[width=\textwidth]{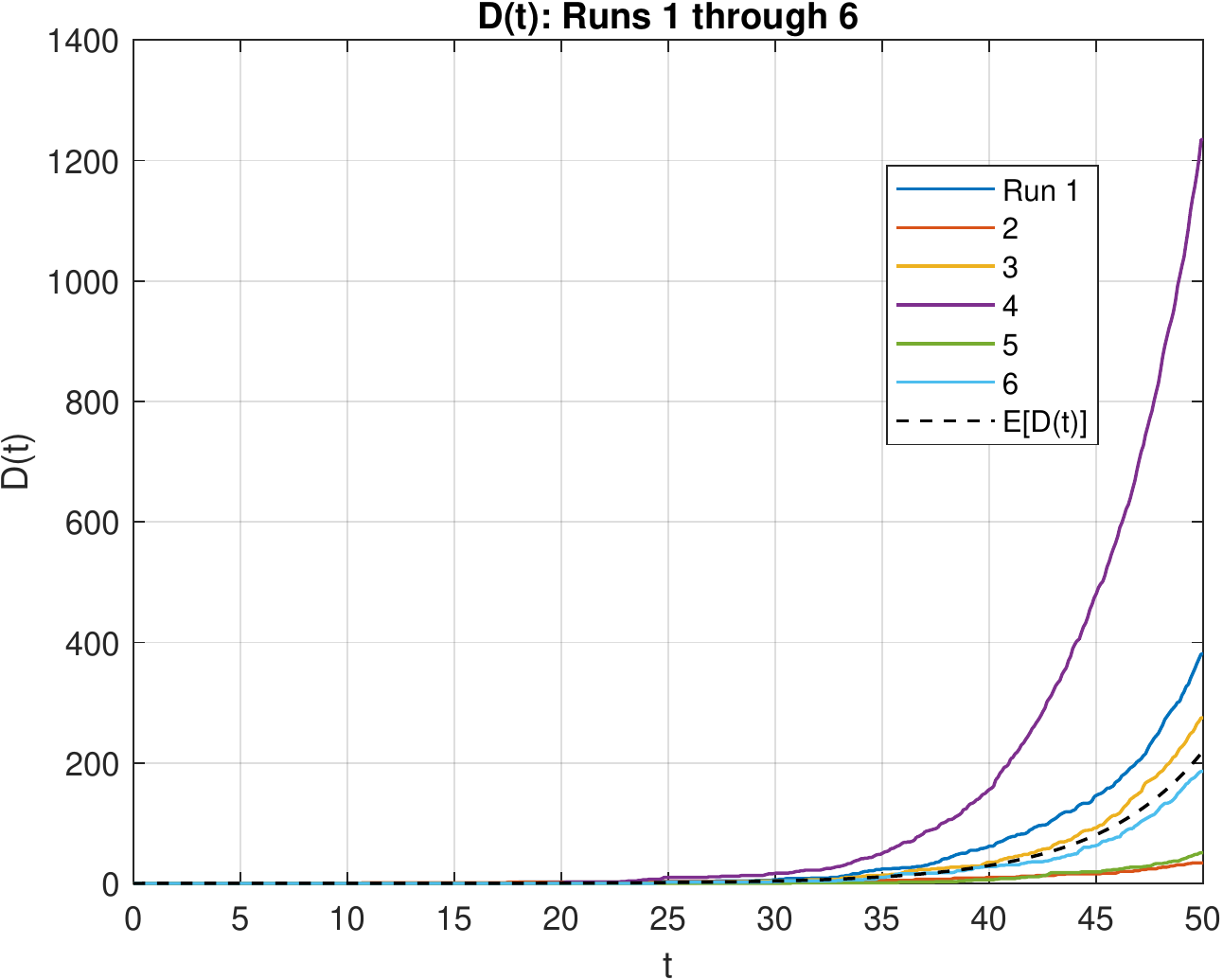}
  \caption{\sf\small The cumulative of deaths $D(t)$: Runs 1-6, the fatality rate (fr) 2\%.}
  \label{fig:D(t)_Runs_1-6}
  \end{minipage}
  \qquad
  \begin{minipage}[t]{0.45\linewidth}
  \centering
  \includegraphics[width=\textwidth]{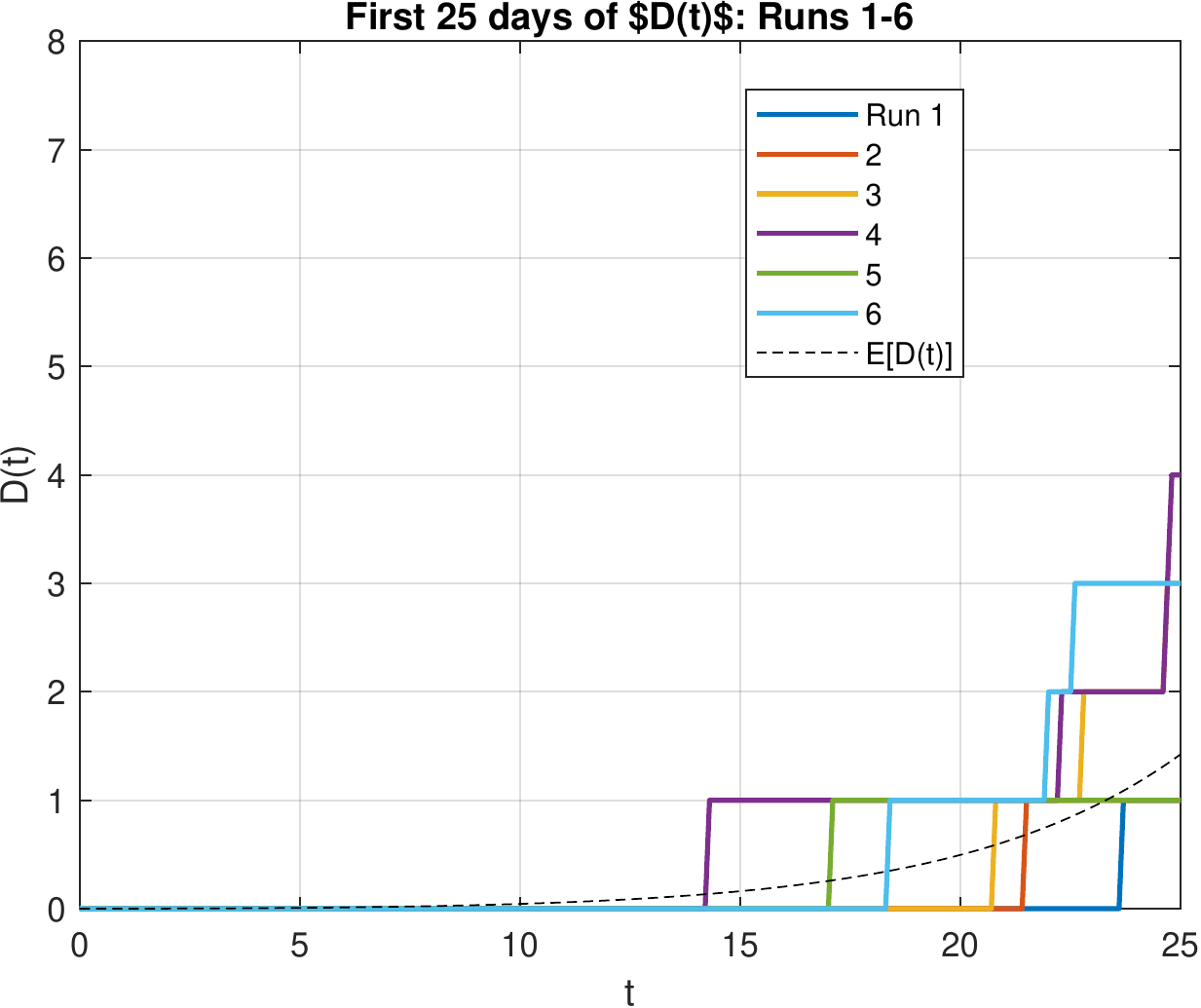}
  \caption{\sf\small The first 25 days of $D(t)$: Runs 1-6.}
  \label{fig:D(t)_Runs_1-6_t=25}
  \end{minipage}
\end{figure}
As an illustration, let us assume fr= 2\% as an overall \emph{fatality rate}.  Simulation of the death process $D(t)$ can be done by randomly splitting the process $R(t)$  with the specified fatality rate and produce $D(t)$ as a sub-process of $R(t)$.  Figures \ref{fig:D(t)_Runs_1-6} shows the result of the six runs.  Figure \ref{fig:D(t)_Runs_1-6_t=25} shows the first 25 days of the same $D(t)$.

\section{The BDI Process $I(t)$ with the Initial Condition $I_0\geq 1$ }  \label{Sec:Initially-nonzero}

By looking at the semi-log plots of simulation runs of the process $I(t)$, some readers may wonder whether the huge variations among different runs may have to do with the fact that some simulation runs remain zero for a considerable period. This question can be answered by examining the process $I(t)$ with the initial condition $I_0=1$. In referring to (\ref{Solution-PGF}), the second term represents the PGF of the BD process without immigration ($\nu$=0), which is often referred to as the \emph{simple birth-and-death process} or as \emph{Feller-Arley (FA) process}. \footnote{In his seminal paper of 1939 \cite{feller:1939}, W. Feller (1906-1970) introduced this process for the study of population growth of some species. N. Arley \cite{arley:1943} applied this model to cascade showers in cosmic ray theory.}
  
\subsection{Analysis of the modified model}
We rewrite (\ref{Solution-PGF}) as
\begin{align}
G_{\scriptscriptstyle BDI:I_0}(z,t)&=G_{\scriptscriptstyle BDI:0}(z,t)G_{\scriptscriptstyle FA:I_0}(z,t),\label{PGF-product}
\end{align}
where $G_{\scriptscriptstyle BDI:0}(z,t)$ is the PGF of the BDI process with the initial condition $I_0=0$, which we have  studied thus far:
\begin{align} 
G_{\scriptscriptstyle BDI:0}(z,t)&=\left(\frac{a}{\lambda  e^{at}-\mu -\lambda(e^{at}-1)z}\right)^r \label{PGF-BDI:0}
\end{align} 
and
\begin{align}
G_{\scriptscriptstyle FA:I_0}(z,t)&=\left(\frac{\mu(e^{at}-1)+(\lambda-\mu e^{at})z}{\lambda e^{at}-\mu -\lambda(e^{at}-1)z}\right)^{I_0}.    \label{PGF-FA}
\end{align}
The product form expression (\ref{PGF-product}) means that the BDI process with nonzero $I_0$ is the sum of the two independent processes, whose PGFs are given by (\ref{PGF-BDI:0}) and (\ref{PGF-FA}):
\begin{align}
I_v&=I_{\scriptscriptstyle BDI:0}(t) + I_{\scriptscriptstyle FA:I_0}(t).\label{sum-I(t)s}
\end{align}

By defining $\alpha(t)$ (see \cite{bailey:1964}\cite{takagi:2007}) by 
\begin{align}
\alpha(t)\triangleq \frac{\mu(e^{at}-1)}{\lambda e^{at}-\mu}=\frac{\mu}{\lambda}\beta(t), \label{def-alpha(t)}
\end{align}
where $\beta(t)$ was earlier defined by (\ref{45-Part-I}), we can rewrite (\ref{PGF-FA}) as
\begin{align}
G_{\scriptscriptstyle FA:I_0}(z,t)=\left(\frac{\alpha(t)+[1-\alpha(t)-\beta(t)]~z}{1-\beta(t)z}\right)^{I_0}.\label{PGF-FA-alpha-beta}
\end{align}
By taking the natural logarithm, differentiating w.r.t. $z$,  and setting $z=1$, we find the expected value of this random process:
\begin{align}
\oI_{\scriptscriptstyle FA:I_0}(t)& \triangleq\Ex[I_{\scriptscriptstyle FA}(t)]
=I_0\frac{(1-\alpha(t))}{(1-\beta(t))}=I_0e^{at}, \label{Mean-I-FA}
\end{align}
Similarly, we find
\begin{align}
\sigma^2_{\scriptscriptstyle FA:I_0}(t)\triangleq \mbox{Var}[I_{\scriptscriptstyle FA:I_0}(t)]
=I_0\frac{(1-\alpha(t))(\alpha(t)+\beta(t))}{(1-\beta(t))^2}=I_0(\lambda+\mu)e^{at}\frac{(e^{at}-1)}{a}. 
\label{Var-I-FA}
\end{align}
which leads to the following expression\footnote{The expression given in our earlier version \cite{kobayashi:2020b} was incorrect.} for the \emph{coefficient of variation} (CV) of the FA process, when $a>0$:
\begin{align}
c_{\scriptscriptstyle FA:I_0}(t)\geq \sqrt{\frac{\lambda+\mu}{aI_0}}, ~~\mbox{for all}~~t.\label{CV-FA}
\end{align}

For the case of our running example with $\lambda=0.3$ and $\mu=0.1$, the RHS of the above becomes $\sqrt{2/I_0}$.
When $I_0=1$, and the CV $c_{\scriptscriptstyle FA:1}(t) \approx 1.414$ for all $t$, which is somewhat larger than the CV of the BDI process with $I_0=0$, which is $c_{\scriptscriptstyle BDI:0}(t)\approx r^{-1/2}=1.225$ (See Part I, p. 19 Example 2). 

\subsubsection{When $I_0=1$:}  

For the case $I_0=1$, it is easy to expand the PGF (\ref{PGF-FA}) in powers of $z^n$, obtaining 
\begin{align}
P_0^{\scriptscriptstyle FA:1}(t)&=\alpha(t)  \label{P_0-FA}\\
P_n^{\scriptscriptstyle FA:1}(t)&=(1-\alpha(t))(1-\beta(t))\beta^{n-1}(t),~~\mbox{for}~~n\geq 1.\label{P_n-FA}
\end{align}

We further analyze the above result depending (i) $a>0$, (ii) $ a<0$,  or (iii) $a=0$.
\begin{itemize}

\item[(i)] When $a=\lambda-\mu>0$:  We find
\begin{align}
\lim_{t\to\infty}\alpha(t)=\frac{\mu}{\lambda}=R_0^{-1}<1, ~~\mbox{and}~~\lim_{t\to\infty}\beta(t)=1,
\end{align}
where  
\begin{align}
R_0=\frac{\lambda}{\mu}>1  \label{def-R_0}
\end{align}
is called the \emph{basic reproduction number} in epidemiology, as defined in (32) of Part I.  The distribution form of (\ref{P_n-FA}) is very similar to (\ref{PMF-I(t)}): it is a geometric distribution of the form $\propto \beta^n(t)$, with $\beta(t)\approx 1$,  and has a long tail similar to Figures 7-12 in Part I. 

\item[(ii)] When $a<0$:  We have
\begin{align}
\lim_{t\to\infty}\alpha(t)=1, ~~\mbox{and}~~\lim_{t\to\infty}\beta(t)=\frac{\lambda}{\mu}=R_0<1,
\end{align}

\item[(iii)] When $a=0$, (i.e., $\mu=\lambda$): We rearrange (\ref{def-alpha(t)}) as
\begin{align}
\alpha(t)=\frac{\mu(e^{at}-1)}{\lambda(e^{at}-1)+a},~~\mbox{and}~~\beta(t)=\frac{\lambda(e^{at}-1)}{\lambda(e^{at}-1)+a}.
\end{align}
Divide both the denominator and the nominator by $a$, and let $a\to 0$.  Using the familiar formula $\lim_{a\to 0}(e^{at}-1)/a=t$, we find
\begin{align}
\alpha(t)=\beta(t)=\frac{\lambda t}{1+\lambda t},~~t\geq 0.
\end{align} 
Rearrange the PGF (\ref{PGF-FA}) in a similar fashion, and we find
\begin{align}
G_{\scriptscriptstyle FA:I_0}(z,t)=\left(\frac{\lambda t +(1-\lambda t)z}{1+\lambda t-\lambda tz}\right)^{I_0},
\end{align}
from which we obtain
\begin{align}
\oI_{\scriptscriptstyle FA:I_0}(t)&\triangleq\Ex[I_{\scriptscriptstyle FA:I_0}(t)]=I_0,\\
\sigma^2_{\scriptscriptstyle FA:I_0}(t)&\triangleq \mbox{Var}[I_{\scriptscriptstyle FA:I_0}(t)]
=2I_0\lambda t,
\end{align}
which could have been directly obtained from (\ref{Mean-I-FA}) and (\ref{Var-I-FA}).
The PMFs of (\ref{P_0-FA}) and (\ref{P_n-FA}) will be further simplified, when $\mu=\lambda$:
\begin{align}
P_0^{\scriptscriptstyle FA:1}(t)&=\frac{\lambda t}{1+\lambda t}  \label{P_0-FA:a=0}\\
P_n^{\scriptscriptstyle FA:1}(t)&=\frac{(\lambda t)^{n-1}}{(1+\lambda t)^{n+1}},~~\mbox{for}~~n\geq 1.\label{P_n-FA:a=0}
\end{align}
\end{itemize} 
\begin{figure}[thb]
  \begin{minipage}[t]{0.45\linewidth}
  \centering
  \includegraphics[width=\textwidth]{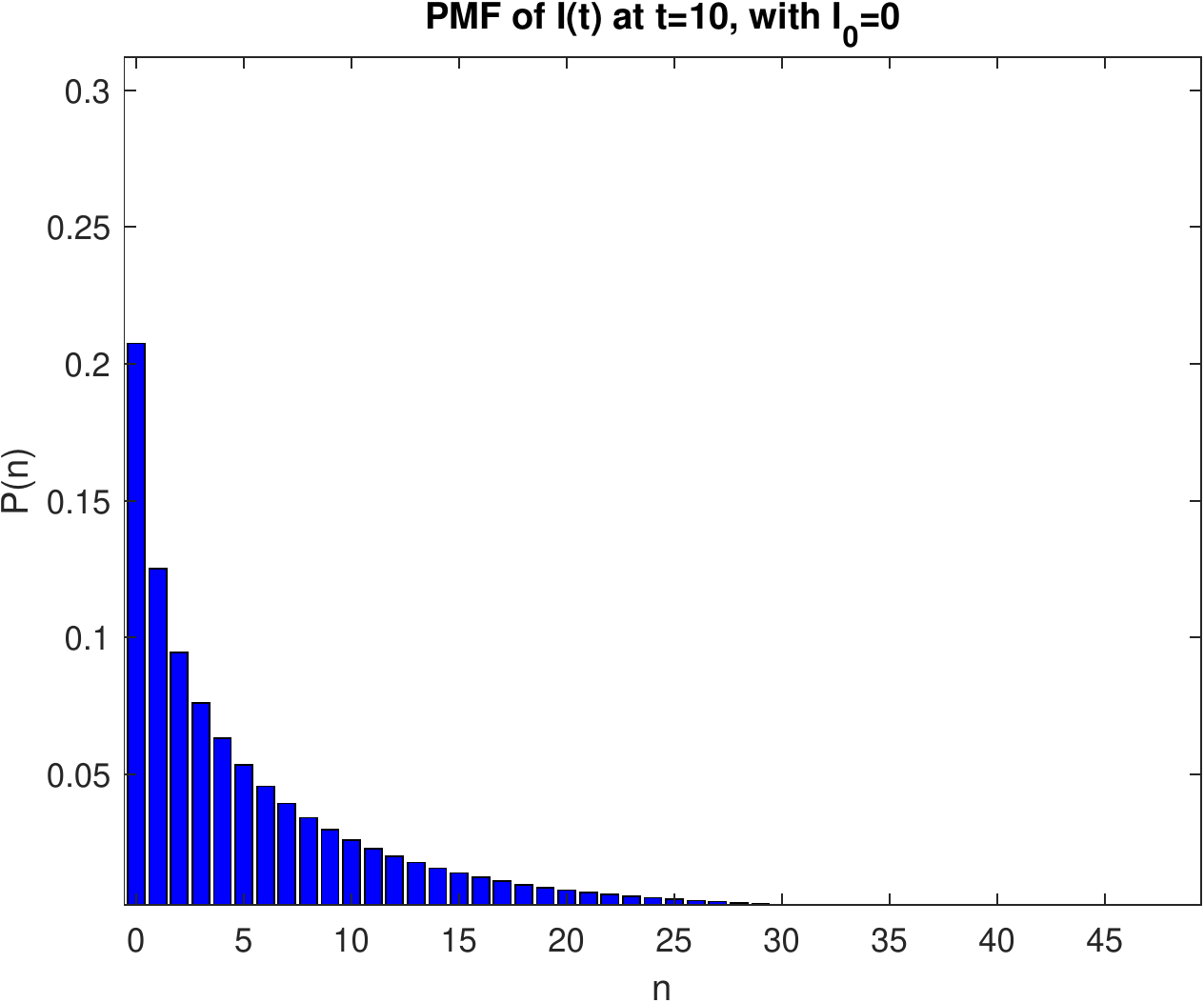}
  \caption{\sf\small PMF of $I(t)$ at $t=10$; $I_0=0$}
  \label{fig:PMF_I(t)_t=10_initial=0}
  \end{minipage}
  \qquad \qquad
  \begin{minipage}[t]{0.45\linewidth}
  \centering
  \includegraphics[width=\textwidth]{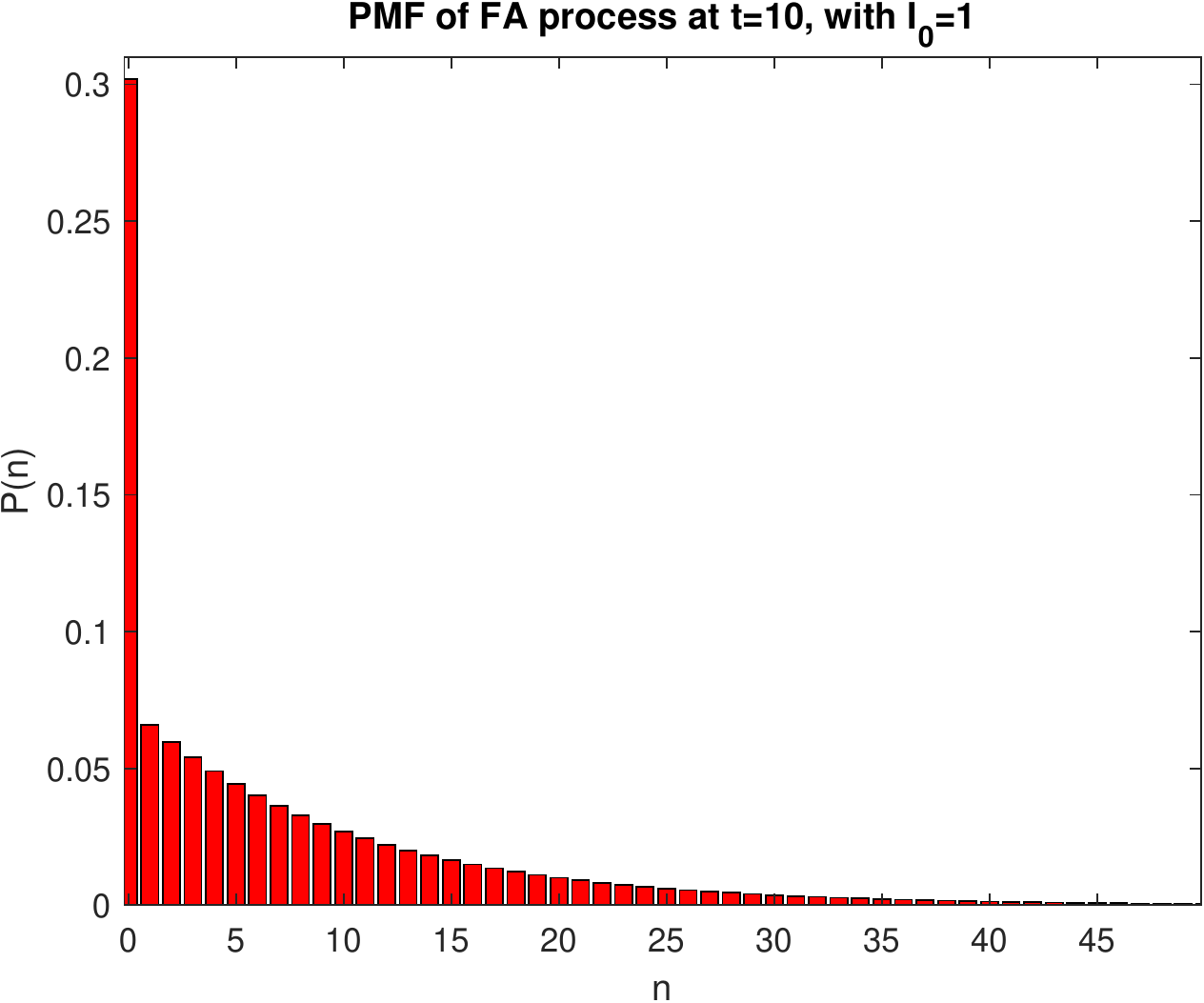}
  \caption{\sf\small PMF of FA Process at $t=10$, $I_0=1$} 
  \label{fig:PMF_FA_t=10_initial=1}
  \end{minipage}
\end{figure}
\begin{figure}[thb]
  \begin{minipage}[t]{0.45\linewidth}
  \centering
  \includegraphics[width=\textwidth]{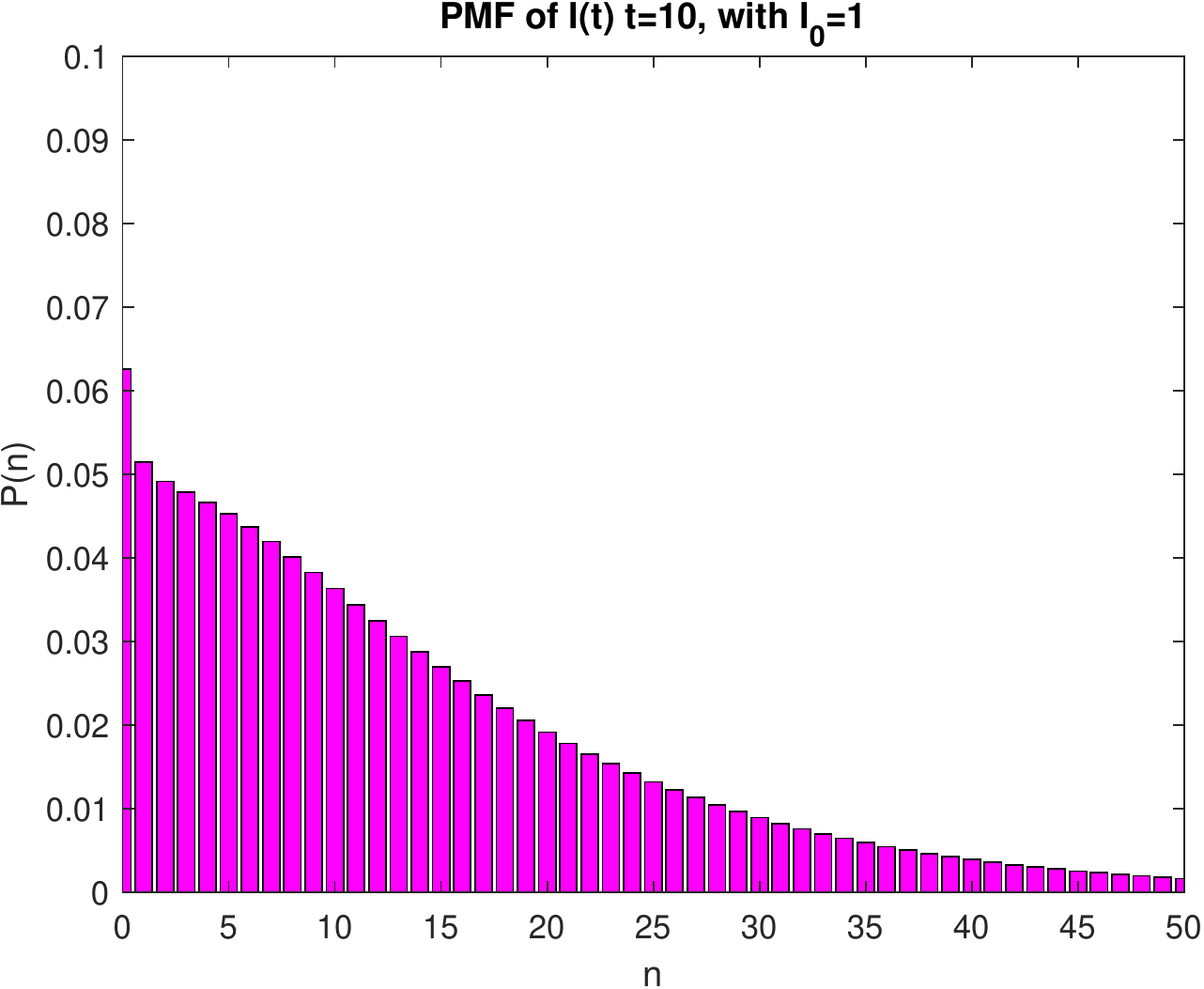}
  \caption{\sf\small PMF of $I(t)$ at $t=10$; $I_0=0$}
  \label{fig:PMF_I(t)_t=10_initial=1}
  \end{minipage}
  \qquad \qquad
  \begin{minipage}[t]{0.45\linewidth}
  \centering
  \includegraphics[width=\textwidth]{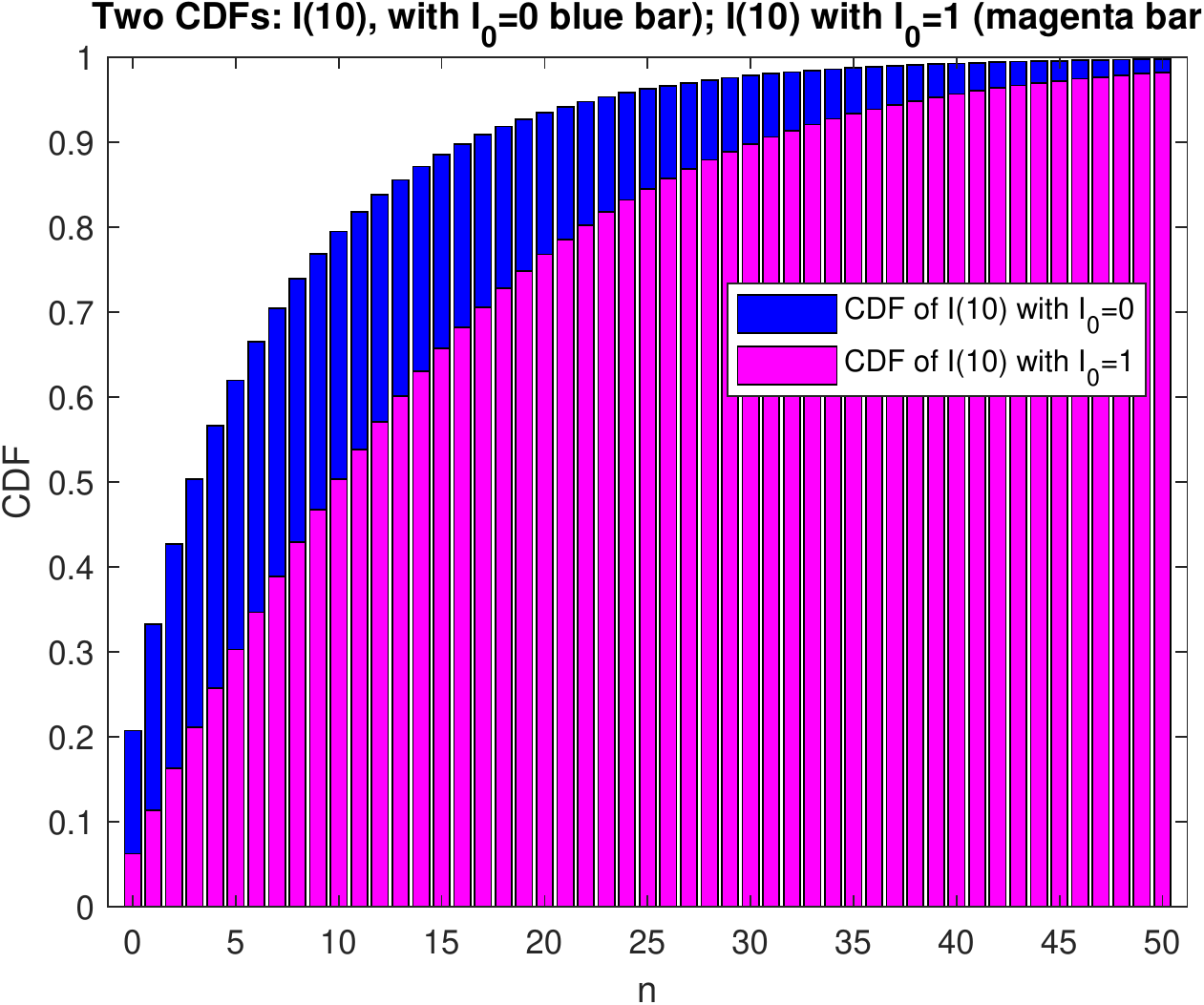}
  \caption{\sf\small CDFs of $I(t)$ at $t=10$,$I_0=0$ (in blue) $I_0=1$ (in Magenta)} 
  \label{fig:TwoCDF_Bars}
  \end{minipage}
\end{figure}

Referring to (\ref{PMF-I(t)}), the PMF of the BDI process with $I_0=1$ is given by the convolution of the two PMFs, for which we have found closed form expressions:
\begin{align}
P_n^{\scriptscriptstyle BDI:1}(t)&=P_n^{\scriptscriptstyle BDI:0}(t)\otimes P_n^{\scriptscriptstyle FA:1}(t)\nonumber\\
%&=\sum_{k=0}^{n-1} {k+r-1\choose k}(1-\beta(t))^r\beta^k(t) (1-\alpha(t))(1-\beta(t))\beta^{n-k-1}(t)
%+ {n+r-1\choose n}(1-\beta(t))^r\beta^n(t)\alpha(t)\nonumber\\
&=(1-\beta(t))^r\left[\sum_{k=0}^{n-1}{k+r-1\choose k}(1-\alpha(t))(1-\beta(t))\beta^{n-1}(t)
+{n+r-1\choose n}\beta^n(t)\alpha(t)\right].  
\label{convolution-PMFs}
\end{align}

In Figure \ref{fig:PMF_I(t)_t=10_initial=0} is the PMF $P_n(t)$ of the $I(t)$ given in (\ref{PMF-I(t)}) at  $t=10$ for our running example, i.e. $a=0.2$ and $r=0.667$ (see Part I, P. 18, Figure 9). Figure \ref{fig:PMF_FA_t=10_initial=1} is the PMF (\ref{P_0-FA}) and (\ref{P_n-FA}).  Figure \ref{fig:TwoCDF_Bars} is the PDF of $I(t)$ with the initial condition $I_0=1$ given by (\ref{PMF-I(t)}).  We computed this by computing the \emph{convolution} of the above two PDFs, i.e., (\ref{convolution-PMFs}). Figure \ref{fig:TwoCDF_Bars} shows the CDFs of $I(t)$ at $t=0$, with $I_0=0$ (in blue) and with $I_0=1$ (in magenta).  

For the CDF $F_X(x)$ of a non-negative random variable $X$, you can show the following simple formula:\footnote{See e.g.,\cite{kobayashi-mark-turin:2012}, pp.73-74.} 
\begin{align}
\int_0^\infty(1-F_X(x))dx=\Ex[X],
\end{align} 
Thus, the white area should be equal to $\oI(t)=\frac{\nu}{a}(e^{at}-1)=e^{2}-1=7.39-1=6.39$ at $t=10$.  Thus, the blue bar area behind the magenta bars should be equal to $I_0e^{at}=e^2=7.39$, an increase in $\Ex[I(t)]$ at $t=10$ due to the presence of an infected person  at $t=0$.

\subsubsection{An efficient computation of the PMF of $I(t)$ with $I_0=1$}
\begin{figure}[bht]
  \begin{minipage}[b]{0.45\linewidth}
  \centering
  \includegraphics[width=\textwidth]{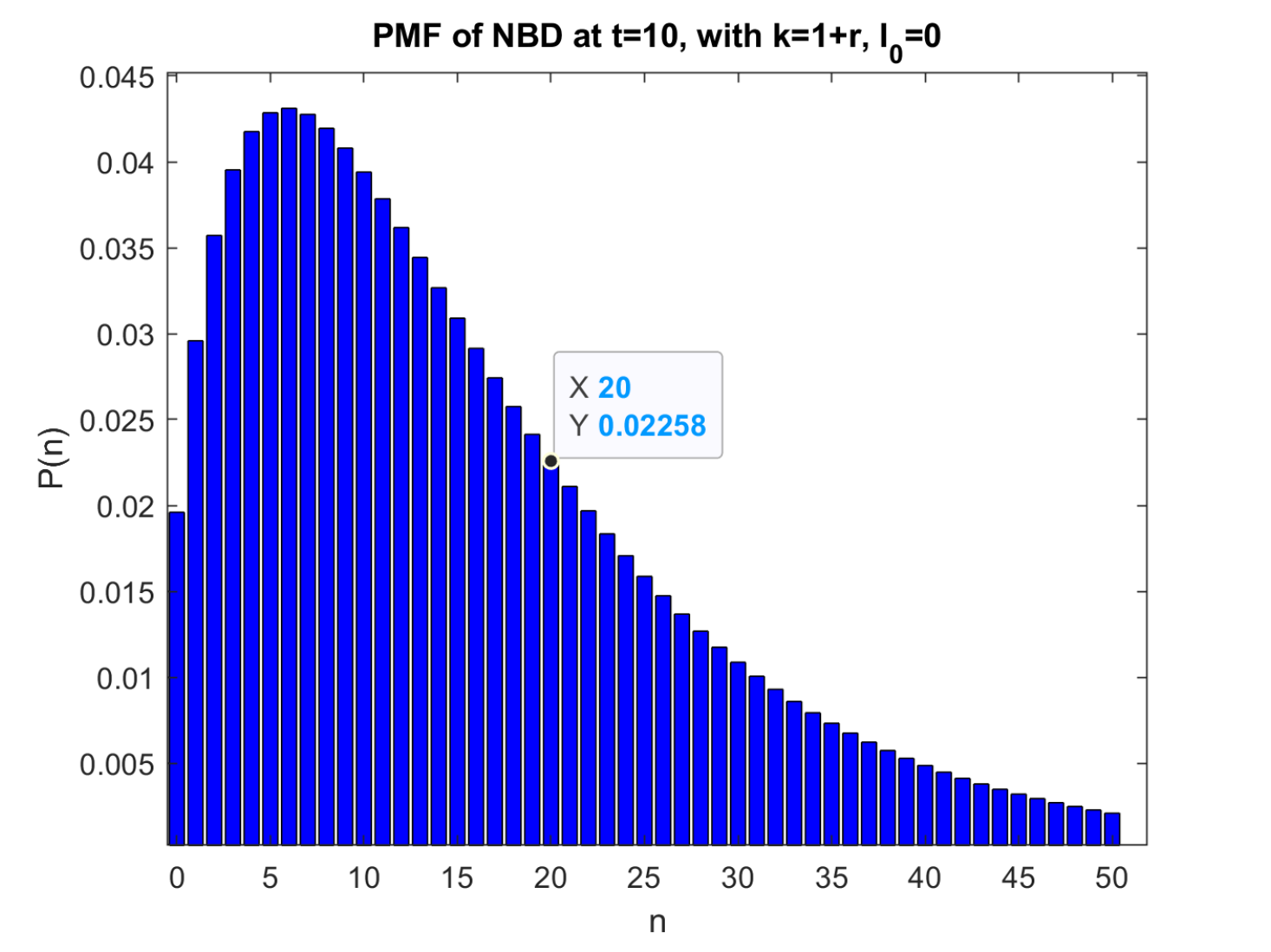}
  \caption{\sf\small NBD with $k=1+r$}
  \label{fig:NBD_k=1+r}
  \end{minipage}
  \qquad \qquad
  \begin{minipage}[b]{0.45\linewidth}
  \centering
  \includegraphics[width=\textwidth]{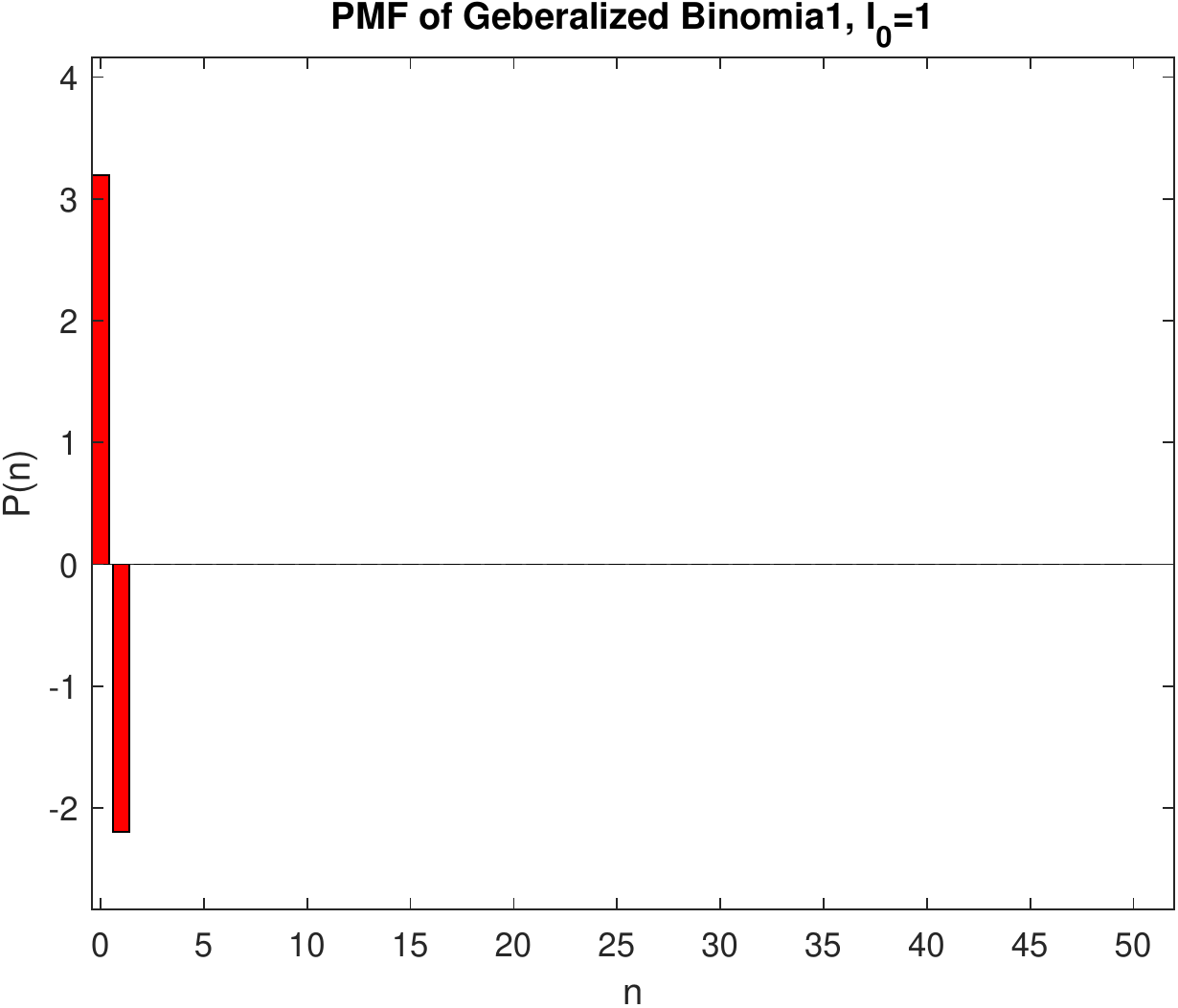}
  \caption{\sf\small Generalized binomial distribution}
  \label{fig:P_0-and-P_1-only}
  \end{minipage}
\end{figure}

An alternative and computationally more efficient way to compute the $P_n^{\scriptscriptstyle BDI:1}$ will be presented below.  We use alternative representation of the PGF (\ref{PGF-FA}) as follow
\begin{align}
G_{\scriptscriptstyle FA:I_0}(z,t)&=\left(\frac{1-\beta(t)}{1-\beta(t)z}\right)^{I_0}
(1-p(t)+p(t)z)^{I_0},
\end{align}  
where
\begin{align}
p(t)=\frac{1-\alpha(t)-\beta(t)}{1-\alpha(t)}=\frac{\lambda-\mu e^{at}}{a}.
\end{align}
Then we can have an alternative product representation to (\ref{PGF-product}), as follows.
\begin{align}
G_{\scriptscriptstyle BDI:I_0}(z,t)&=\left(\frac{1-\beta(t)}{1-\beta(t)z}\right)^{I_0+r}
(1-p(t)+p(t)z)^{I_0}. \label{PGF-alternative-product}
\end{align}
We may term the second term in the above product representation as the PGF of a \emph{generalized binomial distribution} in the sense that the coefficients of $z^n$ terms are not necessarily non-negative. For $I_0=1$, this term is extremely simple, having only two terms:  $P_0(t)=1-p(t)$ and $P_1(t)=p(t),$ which makes the convolution extremely simple:
Fig \ref{fig:NBD_k=1+r} and Figure \ref{fig:P_0-and-P_1-only} show the PDFs of the above two PGFs.  Their convolution results in the PMF as shown in Figure \ref{fig:PMF_I(t)_t=10_initial=1}.

\subsection{Simulation results}                                                                   
\begin{figure}[thb]
 \begin{minipage}[t]{0.45\textwidth}
  \centering
  \includegraphics[width=\textwidth]{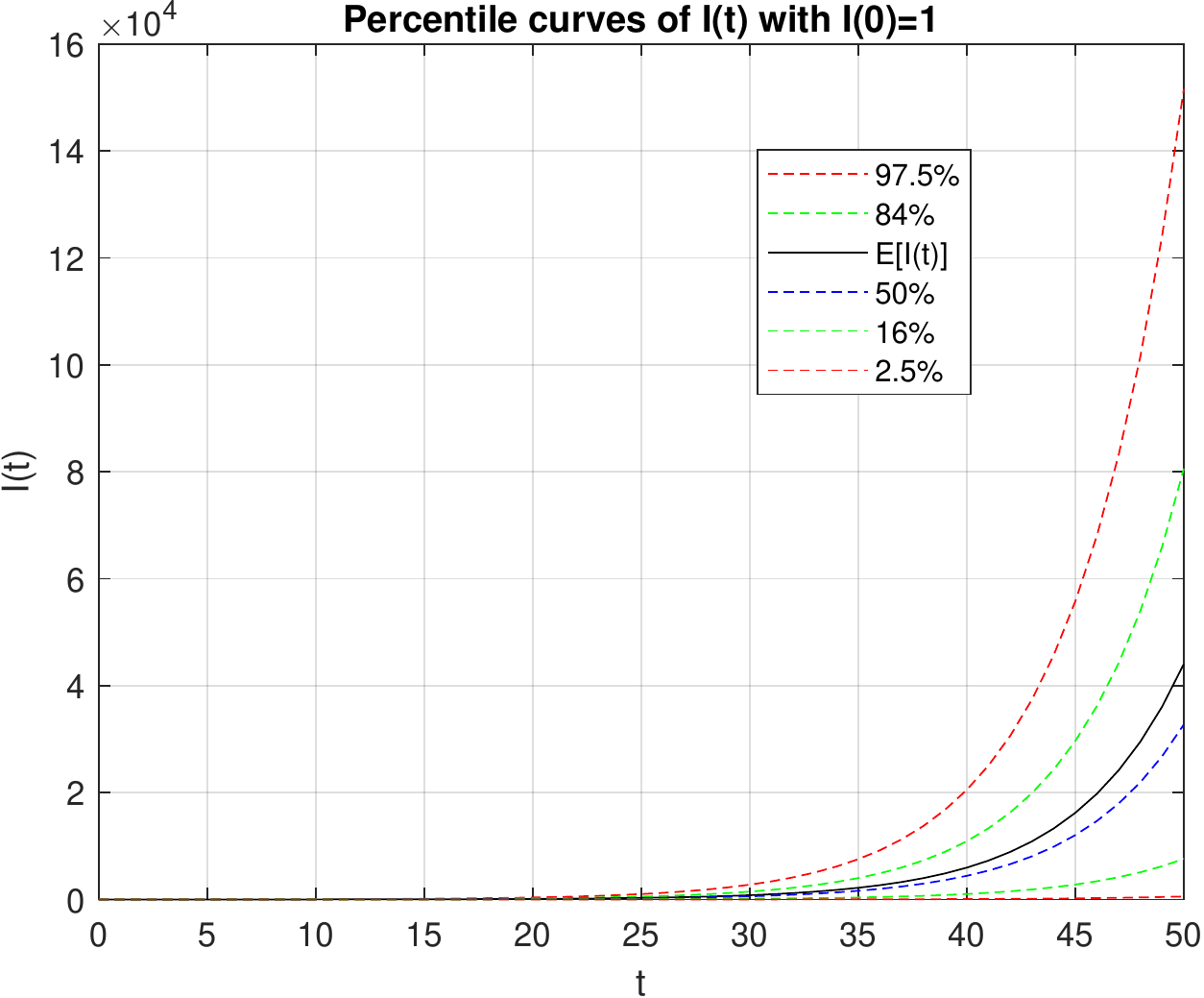}
  \caption{\sf\small Percentile curves of $P_n(t)$ of $I(t)$ with $I(0)=1$; $\lambda=0.3, \mu=0.1, \nu=0.2.$}
  \label{fig:Percentile_curves_I(0)=1_t=50}
  \end{minipage}
  \qquad 
  \begin{minipage}[t]{0.45\textwidth}
  \centering
  \includegraphics[width=\textwidth]{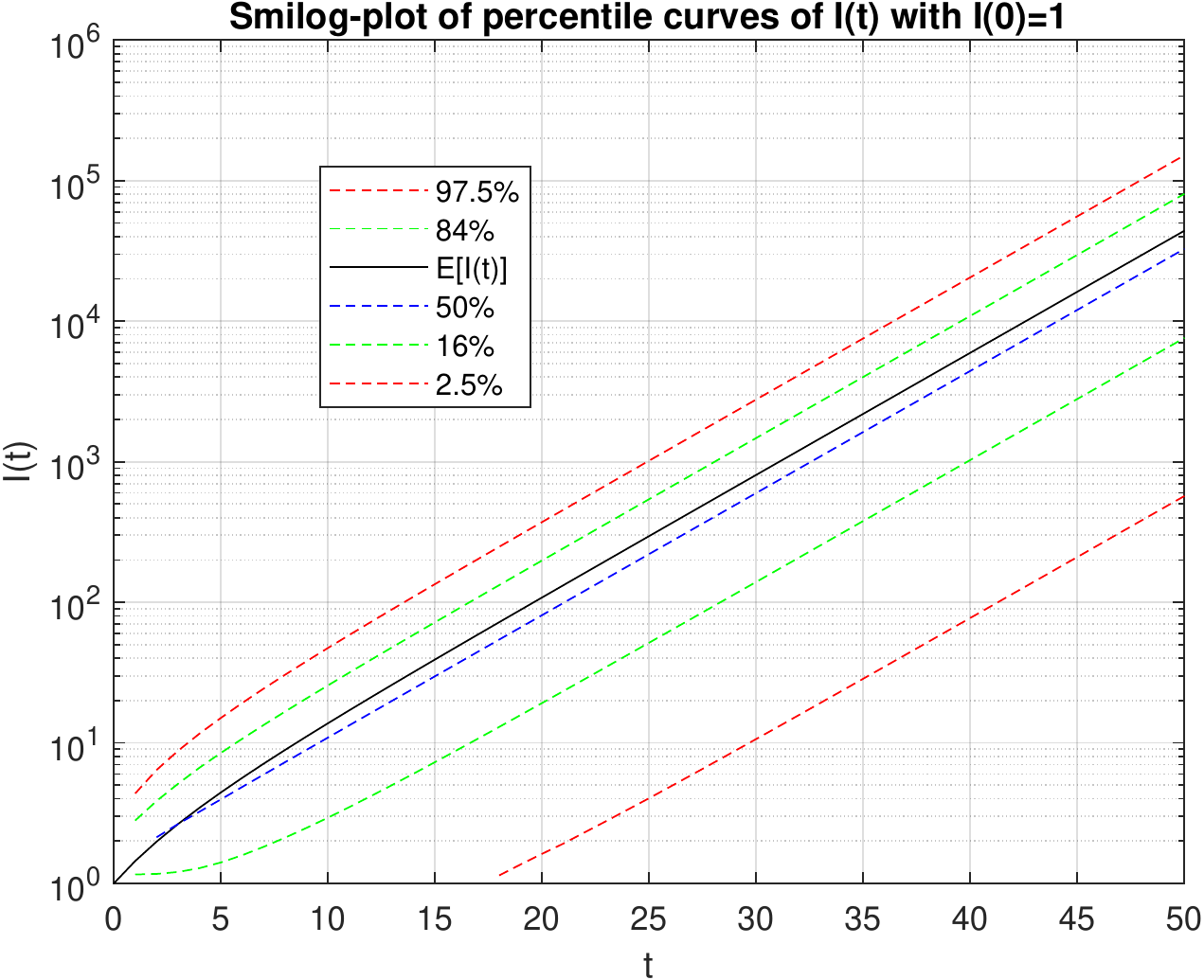}
  \caption{\sf\small Semi-log plot of the percentile curves for $I(t)$ with $I(0)=1$}
  \label{fig:Semi-log_percentile_curves_I(0)=1_t=50}
  \end{minipage}
\hfill
  \begin{minipage}[t]{0.45\linewidth}
  \centering
  \includegraphics[width=\textwidth]{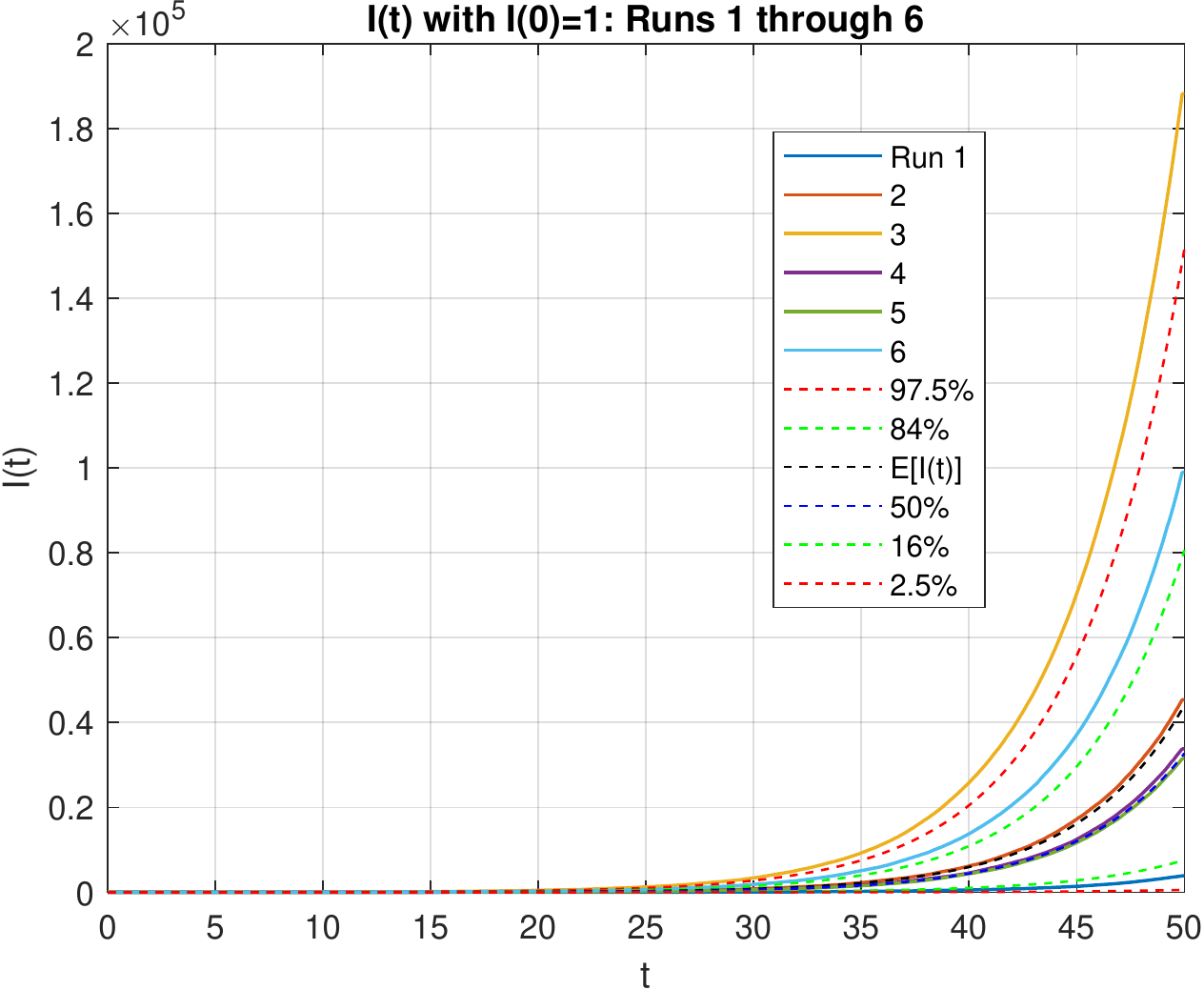}
  \caption{\sf\small The $I(t)$ process with $I(0)=1$, Runs 1-6; $\lambda=0.3, \mu=0.1, \nu=0.2.$}
  \label{fig:I(t)_Runs_1-6_I(0)=1_50}
  \end{minipage}
  \qquad 
  \begin{minipage}[t]{0.45\linewidth}
  \centering
  \includegraphics[width=\textwidth]{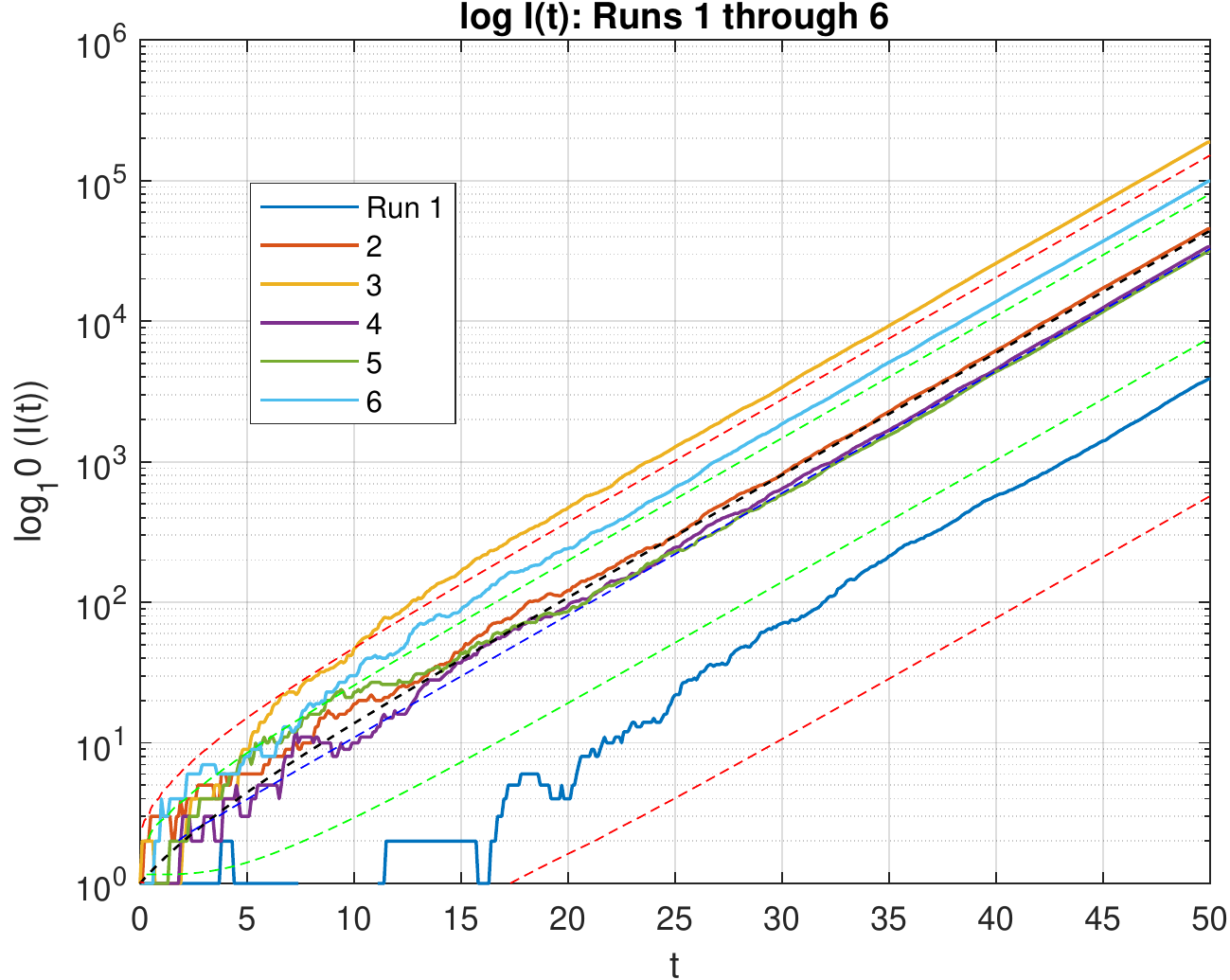}
  \caption{\sf\small Semi-log plots of $I(t)$  with $I(0)=1$, Runs 1-6; $\lambda=0.3, \mu=0.1, \nu=0.2.$}
  \label{fig:Log_I(t)_Runs_1-6_I(0)=1_50}
  \end{minipage}
\end{figure}

The percentile curves for the PMF $P_n(t)$ of $I(t)$ with the initial condition $I(0)=1$ are plotted together with the mean $\oI(t)\triangleq \Ex[I(t)]$ are shown in Figure \ref{fig:Percentile_curves_I(0)=1_t=50} and its semi-log plots are given in Figure \ref{fig:Semi-log_percentile_curves_I(0)=1_t=50}.  

\begin{figure}[thb]
  \begin{minipage}[t]{0.45\linewidth}
  \centering
  \includegraphics[width=\textwidth]{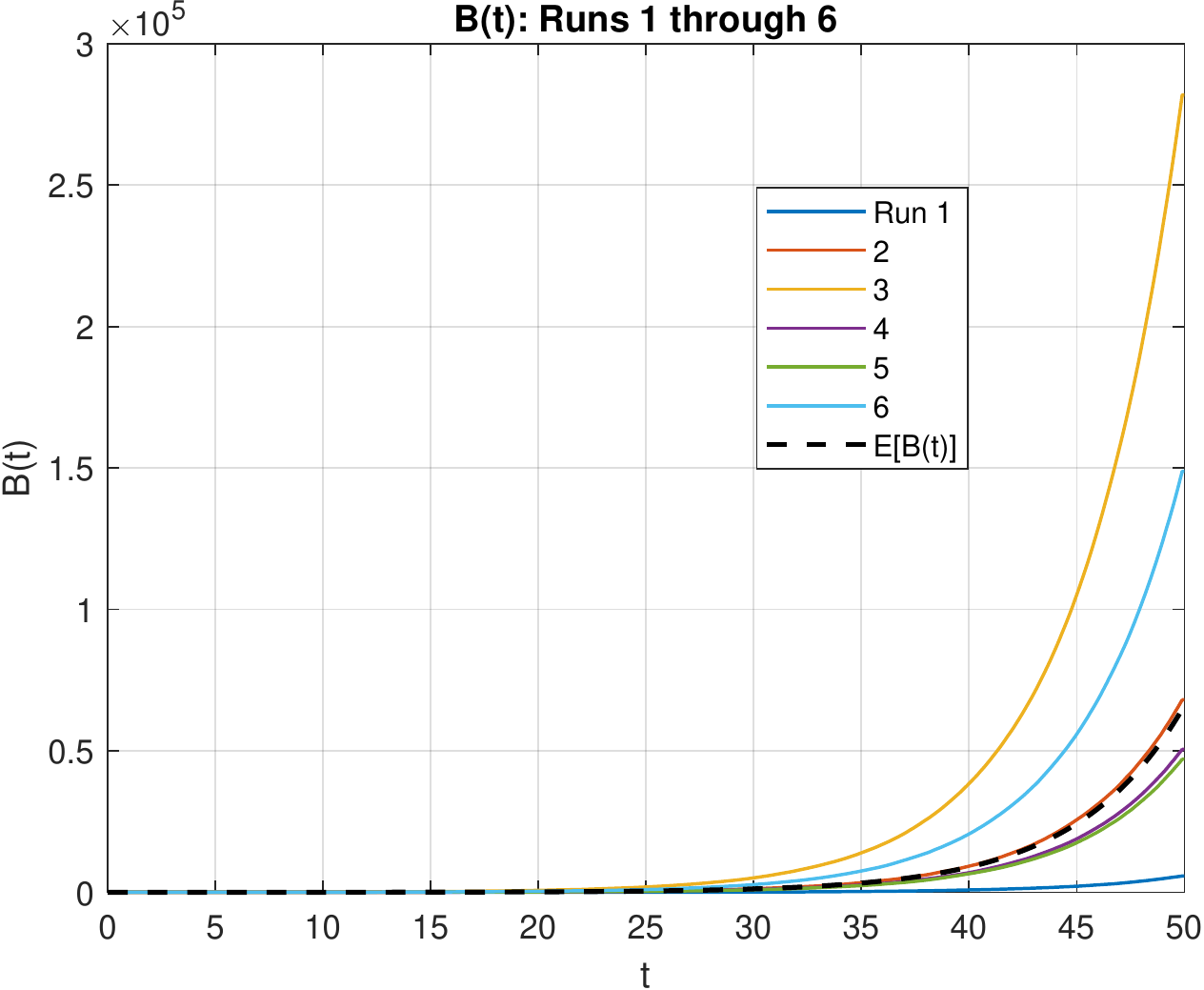}
  \caption{\sf\small The process $B(t)$ with $I(0)=1$: Runs 1-6.}
  \label{fig:B(t)_Runs_1-6_I(0)=1_50}
  \end{minipage}
  \qquad 
  \begin{minipage}[t]{0.45\linewidth}
  \centering
  \includegraphics[width=\textwidth]{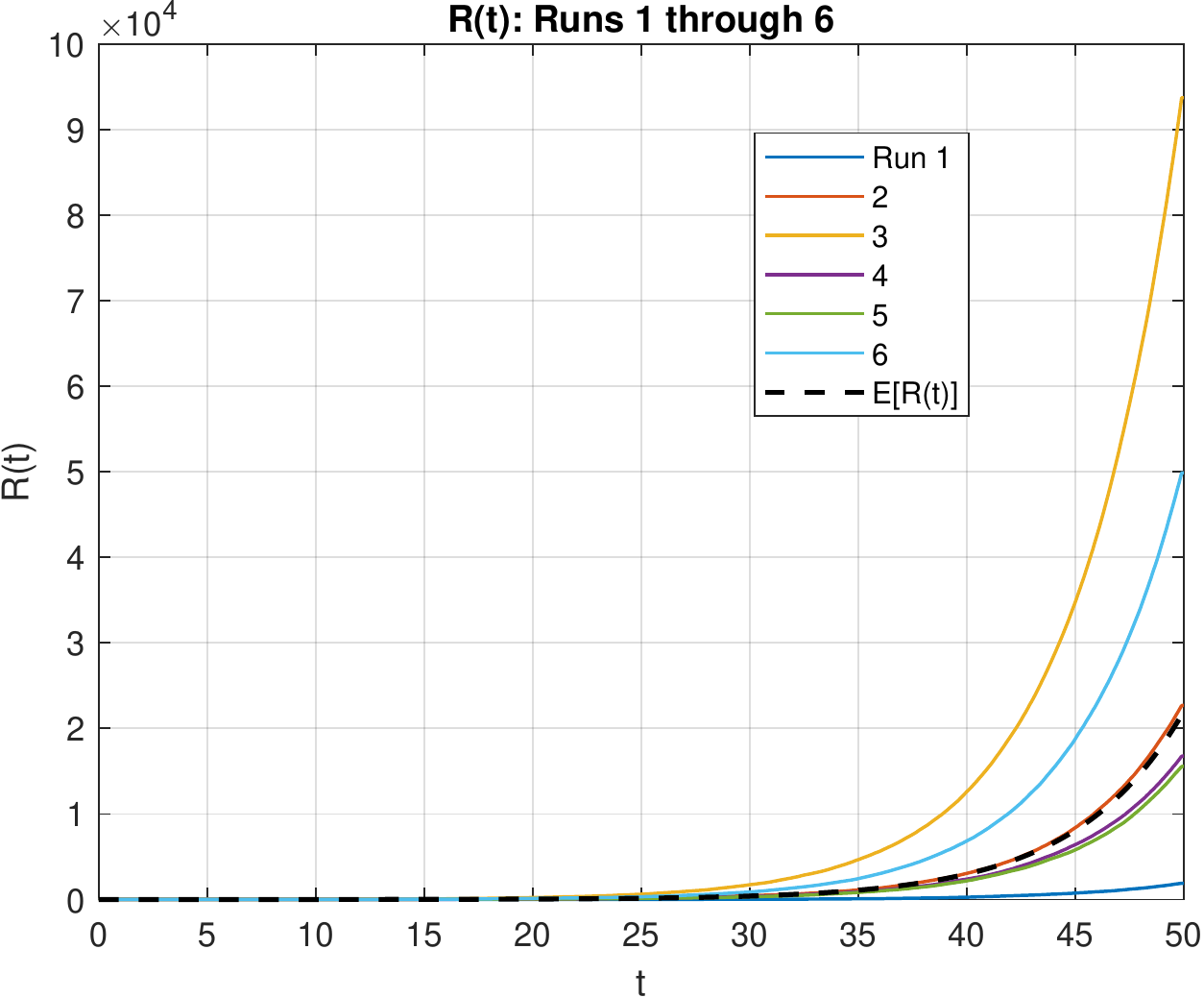}
  \caption{\sf\small The process $R(t)$: Runs 1-6.}
  \label{fig:R(t)_Runs_1-6_I(0)=1_50}
  \end{minipage}
 \end{figure}

In Figure \ref{fig:I(t)_Runs_1-6_I(0)=1_50} we show six simulation runs with the initial condition $I(0)=1$ and their semi-log plots in Figure \ref{fig:Log_I(t)_Runs_1-6_I(0)=1_50}. The large variances among the sample paths still persist, thus we do not see any fundamental differences from the case where the simulations start with the initial condition $I(0)=0$.  This conclusion is not unexpected, if we go back to Equation (\ref{sum-I(t)s}):
\begin{align}
I_{\scriptscriptstyle BDI:I_0}(t)&=I_{\scriptscriptstyle BDI:0}(t) + I_{\scriptscriptstyle FA:I_0}(t).\label{sum-I(t)s-copy},
\end{align} 
in which we set $I_0=1$.  Since the two processes are statistically independent, the sum of their variances is the variance of the summed process:
\begin{align}
\sigma^2_{\scriptscriptstyle BDI:1}(t)&=\sigma^2_{\scriptscriptstyle BDI:0}(t)
+ \sigma^2_{\scriptscriptstyle FA:1}(t),
\end{align}
where the first term on RHS is from Part I (53)
\begin{align}
\sigma^2_{\scriptscriptstyle BDI:0}(t)&=\frac{\nu(\lambda e^{at}-\mu)(e^{at}-1)}{a^2}
\approx\frac{\nu\lambda}{a^2}e^{2at},
\end{align}
and the second term is from (\ref{Var-I-FA})
\begin{align}
\sigma^2_{\scriptscriptstyle FA:1}(t)&=(\lambda+\mu)e^{at}\frac{(e^{at}-1)}{a}\approx\frac{(\lambda+\mu)}{a}e^{2at},
\end{align}
which leads to
\begin{align}
\sigma^2_{\scriptscriptstyle BDI:1}(t)&\approx \left(\frac{\nu\lambda}{a^2}+\frac{(\lambda+\mu)}{a}\right)e^{2at}
=\frac{(\nu+\lambda)\lambda-\mu^2}{(\lambda-\mu)^2}e^{2at}.
\end{align}
The expected value of the process $I(t)$ with $I(0)=1$ is from (\ref{Mean-I-FA}) 
\begin{align}
\Ex[I_{\scriptscriptstyle BDI:1}(t)]&=(1+\frac{\nu}{a})e^{at}-\frac{\nu}{a}\approx\frac{\nu+\lambda-\mu}{\lambda-\mu}e^{at},
\end{align}
Thus, the CV of the the process $I(t)$ remains constant for all $t$ whether not the initial condition is zero or nonzero
\begin{align}
c_{\scriptscriptstyle BDI:1}(t)=\frac{\sqrt{(\nu+\lambda)\lambda-\mu^2}}{\lambda-\mu}= 1.87,
\end{align}
which is even larger than the CV of $I(t)$ with $I(0)=0$, which is 1.225.

\section{Concluding Remarks and Future Plans}

In the present report, we have presented simulation results by implementing a simulator of our stochastic model of an epidemic disease analyzed in Part I \cite{kobayashi:2020a}.  The simulated infection process $I(t)$  indeed exhibits huge variations among different simulation runs. In this report, however, we presented only six consecutive runs in the interest of space. By presenting results of many more runs we could certainly show an even larger disparity between the worst and the best scenarios of a given BDI process $I(t)$. By showing only several runs, however, we can sufficiently demonstrate the great value of a stochastic model in that a deterministic model is more often than not far from any sample path.  In the same token, a single sample path of $I(t)$, which we can observe in a real situation, can contain limited information about the ensemble of the process $I(t)$.  

In the current pandemic of COVID-19, most experts and policy makers in Japan seem looking for all possible factors that might help them understand why they see such enormous disparities between Japan and many countries in the world in terms of the magnitude of infected populations and death tolls. Our analysis and simulation shows that the epidemic process which is a branching process and is driven by inherently positive feedback can have enormous variations in the initial build-up phase, by mere luck or probabilistic chances, sometimes by factor of 100 or more, among environments with identical conditions, i.e., environments with the same $\lambda, \mu$ and $\nu$ parameters in our BDI process based model.

In both analysis and simulations we have so far assumed a time-homogeneous model, whether the model parameters do not change in time.
In Part III-A\cite{kobayashi:2021a}, we will report on our analysis of time-nonhomogeneous models, whereby we allow the model parameters $\lambda(t)$ and $\mu(t)$ to be arbitrary functions of time.  This generalization will help us  better understand, for instance, how a change in \emph{social distancing}, availability of effective vaccines  and/or an improvement/degradation in medical treatment will affect the infection process. In Part III-B \cite{kobayashi:2021b} we will report simulation results to help the reader better understand the significance of the analysis of the time-nonhomogeneous models and augment the analysis, which is limited in finding a closed form solution for a full-fledged BDI process.

In Part IV \cite{kobayashi:2021c}, we plan to develop a statistical theory to estimate the model parameters from real data of COVID-19 epidemics and demonstrate  how our stochastic model can be used in predicting the future behavior of an infectious disease, given its past and present value.  
  
Part V \cite{kobayashi:2021d} will be devoted to use of saddle-point integration technique to approximately characterize the internal infection process $B(t)$ and the recovery process $R(t)$, which seem to defy an exact solution unlike the process $I(t)$ for which we have an exact time-dependent probability distribution function.

Our BDI process based model has a fundamental advantage over nonlinear models such as SIR model and other models dominantly used in the epidemiology community in that our model should be readily extensible to more complex and realistic situations where different \emph{variants} of the infectious diseases may coexist, and different \emph{classes} of susceptible populations (e.g., aged population, those with comorbidity, etc.).  These situations can be modeled by introducing a set of model parameters $\lambda_r, \mu_r$ and $\nu_r$, where $r=(v, c)$ represent the variant $v$ and the class $c$.   
, 
\section*{Acknowledgments}
\addcontentsline{toc}{section}{Acknowledgments}
I thank Prof. Brian L. Mark of George Mason University for his valuable suggestions and help in the MATLAB simulation.  Were it not for his kind help, I could not have completed this rather laborious study.

\bibliographystyle{ieeetr}
\bibliography{infections}

\end{document}